\documentclass[showpacs,prd,twocolumn,floatfix,
nofootinbib,superscriptaddress]{revtex4-1}
\usepackage{graphicx}

\newcommand{\be}{\begin{equation}}
\newcommand{\ee}{\end{equation}}
\newcommand{\nl}{\nonumber \\}

\newcommand{\psibar}{\overline{\Psi}}

\begin{document}

\title{$B \rightarrow D l \nu$ form factors at nonzero recoil and extraction 
of $|V_{cb}|$}

\author{Heechang Na} 
\affiliation{Department of Physics and Astronomy, University of Utah, 
Salt Lake City, Utah 84112, USA}
\author{Chris M. Bouchard}
\affiliation{Physics Department, College of William and Mary, Williamsburg,
 Virginia 23187, USA}
\author{{G. Peter} Lepage}
\affiliation{Laboratory of Elementary Particle Physics,
Cornell University, Ithaca, NY 14853, USA}
\author{Chris Monahan}
\affiliation{Department of Physics and Astronomy, University of Utah, 
Salt Lake City, Utah 84112, USA}
\author{Junko Shigemitsu}
\affiliation{Department of Physics,
The Ohio State University, Columbus, OH 43210, USA}

\collaboration{HPQCD Collaboration}


\begin{abstract}
We present a lattice QCD calculation of the $B \rightarrow D l \nu$ 
semileptonic decay form factors $f_+(q^2)$ and $f_0(q^2)$ for the 
entire physical $q^2$ range. Non-relativistic QCD (NRQCD) bottom quarks 
and Highly Improved Staggered Quark (HISQ) charm and light quarks are 
employed together with $N_f = 2+1$ MILC gauge configurations. 
A joint fit to our lattice and BABAR experimental data allows an 
extraction of the CKM matrix element $|V_{cb}|$. We also determine the 
phenomenologically interesting ratio $R(D) = {\cal B}(B \rightarrow 
D \tau \nu_\tau) / {\cal B}(B \rightarrow D l \nu_l)$ ($l = e, \mu$).  
We find $|V_{cb}|_{excl.}^{B \rightarrow D} = 0.0402(17)(13)$, where the first error
 consists of the lattice simulation errors and the experimental statistical error 
and the second error is the experimental systematic error.
  For the branching fraction ratio we find $R(D) = 0.300(8)$.

\end{abstract}


\maketitle


\section{Introduction}
Studies of the heavy-to-heavy semileptonic decays, $B \rightarrow D l \nu$ 
and $B_s \rightarrow D_s l \nu$,  lead to a wealth of interesting 
and important physics.   These decays can be used, for example, to 
extract the Cabibbo-Kobayashi-Maskawa matrix element $|V_{cb}|$, 
providing an independent check on previous determinations coming from 
$B \rightarrow D^* l \nu$ decays.  There is currently a 
$\sim \,$3$\;\sigma$ tension between the exclusive $|V_{cb}|$ based on $B \rightarrow D^* l \nu$ 
decays at zero recoil and inclusive $|V_{cb}|$ determinations \cite{pdg}. A recent update \cite{DstarMILC} by 
the Fermilab Lattice and MILC collaborations finds 
$|V_{cb}|^{B \rightarrow D^*}_{excl.} = 0.03904 (49)_{expt.}
 (53)_{QCD} (19)_{QED}$, 
whereas the most accurate analysis of inclusive semileptonic decays 
\cite{incvcb} 
 gives $|V_{cb}|_{incl.} = 0.04221 (78) $. 
The current uncertainty in $|V_{cb}|$ leads to the 
dominant error in several important Standard Model predictions for 
rare decays, such as $B_s \rightarrow \mu^+ \mu^-$, $K \rightarrow \pi 
\nu \overline{\nu}$, as well as for the CP violation parameter 
$\epsilon_K$.  Reducing this uncertainty will have an impact 
on precision Flavor Physics.

\vspace{.1in}
In order to get more insight into the tension between inclusive and 
exclusive $|V_{cb}|$ 
 it is crucial to determine  
$|V_{cb}|_{excl.}$ using  channels other than $B \rightarrow D^* l \nu$  and 
also by considering the entire physical $q^2$ range, rather than just the zero 
recoil point. There has been considerable progress on this front.  
A very recent paper by the Fermilab Lattice and MILC
 collaborations, using heavy clover bottom and charm quarks,
 finds $|V_{cb}|_{excl.}^{B \rightarrow D} = (39.6 \pm 1.7_{QCD + exp.} 
\pm 0.2_{QED}) \times 10^{-3}$ from 
$B \rightarrow D l \nu$ lattice form factors and BABAR data \cite{DMILC}. 
And in the present article we give new results on $B \rightarrow D l \nu$ 
form factors 
based on the non-relativistic QCD (NRQCD) action for bottom and 
the Highly Improved Staggered Quark (HISQ) action for charm quarks.
We also combine our lattice form factor results with 
BABAR data to extract
\be
\label{vcb1}
|V_{cb}|_{excl.}^{B \rightarrow D} = 0.0402 (17)(13),
\ee
where the first error comes from the lattice simulation errors and the statistical error
 from  experiment, and the second error is the systematic error from experiment. 

\vspace{.1in}
Interesting physics may also reside in the ratio $R(D) = 
{\cal B}(B \rightarrow D \tau \nu_\tau) / 
{\cal B}(B \rightarrow D l \nu_l)$ ($l = \mu$ or $e$).  BABAR has reported \cite{babar1}
an excess in this ratio over  Standard Model expectations.  The $\tau$ lepton is 
considerably heavier than the electron or the muon, which means that the branching fraction 
into $\tau, \nu_\tau$ is sensitive to both the vector form factor $f_+(q^2)$ and 
the scalar form factor $f_0(q^2)$, while the latter does not contribute 
for decays into $\mu , \nu_\mu$ or $e, \nu_e$. This could allow 
 scalar contributions from new physics 
to enter just in the numerator of $R(D)$
 and thereby explain the apparent excess.
 In order to confirm or reject the $R(D)$ anomaly as a 
true new physics effect, it is important to scrutinize the current 
 Standard Model prediction for $R(D)$.  Ref. \cite{RdMILC} gave the first unquenched 
lattice result 
for $R(D)$. 
Using the new form factors presented in this article we find,
\be
\label{rd0}
R(D) = 0.300(8),
\ee
the most accurate Standard Model prediction to date. 
 
\vspace{.1in}
The rest of this article is organized as follows. 
 Sec. II gives details
 of the lattice setup 
for this project, introduces the relevant bottom-charm currents and 
defines the vector and scalar form factors $f_+(q^2)$ and $f_0(q^2)$. 
  Sec. III introduces the two- and three-point correlators 
that we simulate and describes our correlator fits and extraction of 
form factors. In Sec. IV we explain how our results for lattice form 
factors are extrapolated to the physical, i.e. chiral/continuum, limit. 
In Sec. V we discuss our form factor results in the physical 
limit and their errors coming from different sources. We also extract 
the ``slope parameter''  $\rho^2$ for $f_+(q^2)$. 
 In Sec. VI we combine our Standard Model theory results 
with experimental measurements of the $B \rightarrow D l \nu$ 
branching fraction to extract a new value for $|V_{cb}|$. Sec. VII 
is devoted to the ratio $R(D)$. We conclude and summarize in 
Sec. VIII. 
  In  Appendix A we provide the relevant  
information needed to reconstruct our form factors, including correlations.
 Appendix B discusses further details and checks on the
 chiral/continuum/kinematic extrapolations. And in Appendix C we list 
the priors and prior widths used in these  extrapolations. 

\section{  Lattice Setup and NRQCD/Heavy-HISQ Currents }

Table I lists the three coarse ($a \approx 0.12$fm) and 
two fine ($a \approx 0.09$fm) MILC $N_f = 2 + 1$ ensembles \cite{latMILC} used in this study, 
together with some further simulation details.  These MILC configurations employ 
the asqtad action to incorporate up, down and strange sea quarks. Compared 
to our recent $B \rightarrow K l^+ l^-$ \cite{btok1, btok2} and $B_s \rightarrow K l \nu$ \cite{bstok} projects 
we have increased statistics by about a factor of two or more.
 For the 
valence bottom quarks we use the NRQCD action described, 
 for instance, in \cite{nrqcd}.  The valence light and charm quarks are 
represented by the HISQ action \cite{HISQ}.  In Table II
we show the values for valence quark masses.  The NRQCD bottom quark mass $a M_b$ was tuned 
in Ref. \cite{fb} to reproduce the spin averaged $\Upsilon$ mass, whereas the HISQ bare mass $a m_c$ 
was tuned to the $\eta_c$ mass (suitably modified to accommodate the lack of 
 annihilation and electromagnetic contributions in our simulations) in \cite{dtok}.  
The valence HISQ light quark mass $a m_l$ was chosen to be close to the 
light asqtad quark mass in the sea.

\begin{table}
\caption{
Simulation details on three ``coarse'' and two ``fine''  $N_f = 2 + 1$ MILC ensembles.
}
\begin{ruledtabular}
\begin{tabular}{cccccc}
Set &  $r_1/a$ & $m_l/m_s$ (sea)   &  $N_{conf}$&
 $N_{tsrc}$ & $L^3 \times N_t$ \\
\hline
C1  & 2.647 & 0.005/0.050   & 2096  &  4 & $24^3 \times 64$ \\
C2  & 2.618 & 0.010/0.050  & 2256   & 2 & $20^3 \times 64$ \\
C3  & 2.644 & 0.020/0.050  & 1200  & 2 & $20^3 \times 64$ \\
\hline
F1  & 3.699 & 0.0062/0.031  & 1896  & 4  & $28^3 \times 96$ \\
F2  & 3.712 & 0.0124/0.031  & 1200  & 4 & $28^3 \times 96$ \\
\end{tabular}
\end{ruledtabular}
\end{table}

\begin{table}
\caption{
Valence quark masses $a M_b$ for NRQCD bottom quarks and 
$a m_l$ and $a m_c$ for HISQ light and charm quarks.  The last 
column gives 
$Z_2^{(0)}(a m_c)$, the tree-level wave function renormalization 
constant for massive (charm) HISQ quarks \cite{matching}.
}
\begin{ruledtabular}
\begin{tabular}{ccccc}
Set & $a M_b$  & $a m_l$ & $a m_c$ & $  Z_2^{(0)}(a m_c) $   \\
\hline
C1  &  2.650 &  0.0070 & 0.6207  & 1.00495618 \\
C2  & 2.688 & 0.0123  &  0.6300   & 1.00524023 \\
C3  & 2.650 & 0.0246 & 0.6235   & 1.00504054\\
\hline
F1  & 1.832 & 0.00674 & 0.4130   & 1.00103879\\
F2  & 1.826  & 0.01350 & 0.4120   & 1.00102902\\
\end{tabular}
\end{ruledtabular}
\end{table}

\vspace{.1in}
To study the process $B \rightarrow D \, l \nu$, one needs to 
evaluate the matrix element of the bottom-charm charged electroweak  current 
between the $B$ and the $D$ states, $\langle D | (V - A)^\mu | B \rangle$. 
Only the vector current $V^\mu$ contributes to the pseudoscalar-to-pseudoscalar
 amplitude and the matrix 
element can be written in terms of two form factors $f_+(q^2)$ and 
$f_0(q^2)$. These depend only on the square of the momentum transfered 
between the $B$ and the $D$ mesons, $q^\mu = p_B^\mu - p_D^\mu$,
\begin{eqnarray}
\label{f0plus}
\langle D(p_D)| V^\mu| B(p_B) \rangle &=&
f_+(q^2) 
\, \left[ p_B^\mu + p_D^\mu - \frac{M_B^2-M_D^2}{q^2}
\, q^\mu \right] \nonumber \\
 &+& f_0(q^2)
 \; \frac{M_B^2 - M_D^2}{q^2} \; q^\mu .
\end{eqnarray}
Intermediate stages of the analysis are simplified by working with the form factors $f_\parallel$ and $f_\perp$, defined by
\be
\label{fpaperp}
\langle D(p_D)| V^\mu| B(p_B) \rangle =
 \sqrt{2 M_B} \,[v^\mu f_\parallel\,
 + \, p^\mu_\perp  f_\perp ] ,
\ee
with
\be
v^\mu = \frac{p_B^\mu}{M_B} , \qquad p_\perp^\mu
=p_D^\mu - (p_D \cdot v) \, v^\mu .
\ee
In the $B$ rest frame (in this article we only consider $B$ mesons 
decaying at rest) the temporal and spatial parts of (\ref{fpaperp}) 
become,
\begin{eqnarray}
\label{fv0}
\langle D| V^0 |B \rangle &=&
 \sqrt{2 M_B} \,       f_\parallel,  \\
\label{fvk}
\langle D| V^k |B \rangle &=&
 \sqrt{2 M_B} \,p^k_D \, f_\perp .
\end{eqnarray}
Hence, one sees that one can separately determine $f_\parallel$ or 
$f_\perp$  simply by looking at either the temporal or spatial 
component of $V^\mu$. 
The conventional form factors $f_+(q^2)$ and $f_0(q^2)$ can then be obtained from
\be
f_+ = \frac{1}{\sqrt{2 M_B}} \, f_\parallel + \frac{1}{\sqrt{2 M_B}} \,
(M_B - E_D) \, f_\perp ,
\ee
\be
f_0 = \frac{\sqrt{2 M_B}}{(M_B^2 - M_D^2)}\, [ (M_B- E_D) f_\parallel
+ (E_D^2 - M_D^2) f_\perp ] ,
\ee
where $E_D$ is the daughter $D$ meson energy in the $B$ rest frame. 
We generate data for four different $D$ meson momenta, 
$\frac{ 2 \pi}{aL}(0,0,0)$, 
$\frac{ 2 \pi}{aL}(1,0,0)$, 
$\frac{ 2 \pi}{aL}(1,1,0)$, and 
$\frac{ 2 \pi}{aL}(1,1,1)$.

\vspace{.1in} 
Our goal is to evaluate the hadronic matrix elements 
$\langle D | V^0|B\rangle$ and 
$\langle D | V^k|B\rangle$ via lattice simulations.  There are three 
steps in the calculation.  First, one must relate the continuum electroweak 
currents,
 $V^0$ and $V^k$, to lattice operators written in terms of the bottom and charm 
quark fields in our lattice actions.  In the second step the matrix 
elements of these lattice current operators must be evaluated numerically 
and the relevant amplitudes, i.e. the matrix elements between the ground state 
$B$ meson and the ground state $D$ meson with appropriate momenta, must 
be extracted.  This will give us, via Eqs.~(\ref{fv0}) and (\ref{fvk}), the 
form factors $f_\parallel$ and $f_\perp$ as functions of the light quark 
mass and the $D$ momentum.  Finally, in step 3 these numerical results
 must be extrapolated to the physical, chiral/continuum, limit.  
In the next two sections we describe steps 2 and 3 in turn. 
Here we consider step 1 and conclude the
section with a brief overview of the bottom-charm currents used in our 
simulations and of how these effective theory currents are matched to those 
in continuum QCD.

 \vspace{.1in}
Given our NRQCD action for bottom quarks and HISQ action for charm, we have, 
through next-to-leading order (NLO) in $1/M$ and lowest order in $\alpha_s$, the two currents
\begin{eqnarray}
\label{j0}
 J^{(0)}_\mu & = & \bar{\psi}_c \,\gamma_\mu \, \Psi_b, \\
\label{j1}
 J^{(1)}_\mu & = & \frac{-1}{2M_b} \bar{\psi}_c
    \,\gamma_\mu\,\mbox{\boldmath$\gamma\!\cdot\!\nabla$} \, \Psi_b.
\end{eqnarray}
Here $\psi_c$ is the HISQ charm quark field 
(in its four component ``naive fermion'' form) and 
$\Psi_b$ the heavy quark field with the upper two components given by the 
two-component NRQCD fields and the lower two components set equal to zero.
We have matched these effective theory currents to $V_\mu$ in full QCD at 
one-loop order through  
 ${\cal O}(\alpha_s, \frac{\Lambda_{QCD}}{M}, 
\frac{\alpha_s}{aM})$. 
Details of the matching of NRQCD/HISQ currents are given in Ref. \cite{matching}.  
The matching is similar to that employed in recent heavy-to-light 
semileptonic decays (i.e. $B \rightarrow K l^+ l^-$ and $B_s \rightarrow K l \nu$).  
However, there is a difference between matching of NRQCD/massless-HISQ 
and NRQCD/massive-HISQ currents. Massive HISQ fermions have a nontrivial 
wave function renormalization $Z_2^{(0)}$ even at tree-level. 
To ensure that matching coefficients scale as  $\{ 1 + {\cal O} (\alpha_s) + .... \}$, we factor out this tree-level rescaling at the outset.
This means the currents in 
(\ref{j0}) and (\ref{j1}) get multiplied by $ \left (Z_2^{(0)} \right )^{-1/2}$.
 After this rescaling, which one sees from Table II is a very small effect, one has
\be
\label{jmu}
\langle V_\mu \rangle_{QCD}  = ( 1 + \alpha_s
 \, 
\rho_\mu)\,\langle J^{(0)}_\mu \rangle  + 
 \, \langle J^{(1),sub}_\mu \rangle, 
\ee
with
\be
\label{jsub}
 J_\mu^{(1),sub} = J_\mu^{(1)} - \alpha_s \,   \zeta_\mu
J_\mu^{(0)} .
\ee
Here $\rho_\mu$ and $\zeta_\mu$ are the one-loop 
matching coefficients tabulated for $\mu = 0$ and $\mu = k$ in \cite{matching} for several
 $aM_b$ and $a m_c$ values.

\section{ Correlators and Fitting Strategies}

In order to extract $\langle D | J_\mu | B \rangle $ (here we use ``$J_\mu$''
to denote either the full expression for the current on the rhs of (\ref{jmu}) or just 
the lowest order term $J^{(0)}_\mu$), we need to calculate 
the $B$ and $D$ meson two-point correlators and the $J_\mu$ three-point correlator.
We use smeared heavy-light bilinears with Coulomb gauge fixed lattices to represent the $B$ meson. 
For instance, we create a meson at time $t_0$ via
\be
\Phi^{\alpha \dagger}_B(\vec{x},t_0) \equiv a^3 \sum_{\vec{x}^\prime}
 \psibar_b(\vec{x}^\prime,t_0)
 \phi^\alpha(\vec{x}^\prime - \vec{x}) \gamma_5 \psi_l(\vec{x},t_0).
\ee
 For the smearing functions, $\phi^\alpha(\vec{x}^\prime - 
\vec{x})$,
 we use a $\delta$-function local smearing ($\alpha = 1$) 
or Gaussian smearings $ \propto e^{-|\vec{x}^\prime - \vec{x}|^2/
(2 r_0^2)}$, normalized to one ($\alpha = 2$).
We then calculate a $2 \times 2$ matrix of zero momentum $B$  meson correlators with all 
combinations of source and sink smearings,
\be
\label{b2pnt}
C_B^{\beta,\alpha}(t,t_0) = \frac{1}{L^3} \sum_{\vec{x}, \vec{y}} 
\langle \Phi_B^\beta(\vec{y},t) \, \Phi^{\alpha \dagger}_B(\vec{x},t_0) \rangle.
\ee
 We use Gaussian widths in lattice units of size 
$r_0/a = 5$ on coarse ensembles and $r_0/a = 7$  on the fine ensembles.  
For the $D$ meson built from HISQ charm and light quarks we use an interpolating 
operator,
\be
\Phi^\dagger_D(\vec{x}, t_0) = a^3 \bar{\psi}_c(\vec{x}, t_0) \gamma_5 \psi_l(\vec{x}, t_0),
\ee
and construct two-point correlation function with momentum $\vec{p}$,
\be
\label{d2pnt}
C_{D}(t,t_0; \vec{p}) = 
\frac{1}{L^3}{\cal N}_{taste}\sum_{\vec{x}, \vec{y}}  e^{i \vec{p} \cdot (\vec{x} - \vec{y})}
 \langle  \Phi_D(\vec{y},t) 
\Phi_D^\dagger(\vec{x},t_0) \rangle.
\ee
The normalisation factor ${\cal N}_{taste} = 1/16$ for four-component naive HISQ 
quarks, and ${\cal N}_{taste} = 1/4$ when employing the one-component version of HISQ. 
As explained in Ref. \cite{heavy_light} there are no ``taste'' related rescaling factors when a nondoubled 
NRQCD heavy quark propagator is part of the loop, such as in   $C_B^{\beta, \alpha}(t,t_0)$ 
above or in the three-point correlator given below.

\vspace{.1in}
The  three-point correlator of $J_\mu$ can be written as
\begin{eqnarray}
\label{j3pnt}
C^\alpha_J(t,t_0,T;\vec{p}) &=& \frac{1}{L^3} \sum_{\vec{x}, \vec{y}, \vec{z}}
e^{i \vec{p} \cdot (\vec{z} - \vec{x})} \nonumber \\
 && \times \langle \Phi_D(\vec{x},t_0 + T)\, J_\mu(\vec{z},t) \,
\Phi^{\alpha \dagger}_B(\vec{y}, t_0) \rangle. \nonumber \\
\end{eqnarray}
The setup for the three-point correlator in (\ref{j3pnt}) is shown in Fig.~1.  The 
$B$ meson is created at time slice $t_0$.  A current insertion at timeslice $t$, 
$t_0 \leq t \leq t_0 + T$, converts the $b$-quark into a $c$-quark.  The resulting 
$D$ meson is annihilated at timeslice $t_0 + T$.  We have accumulated simulation 
data for four values of $T$: 12, 13, 14, and 15 on coarse and 21, 22, 23, and 24 on fine lattices.
 The source time $t_0$ is picked 
randomly for each gauge configuration
  in order to reduce autocorrelations.  Using translational 
invariance, all data are shifted to $t_0 = 0$ before taking averages and/or doing fits. 
The spatial sums at the source, $\sum_{\vec{x}}$, in Eqs.~(\ref{b2pnt}), 
(\ref{d2pnt}) and (\ref{j3pnt}) are implemented using $U(1)$ random wall sources 
$\xi(x^\prime)$ and $\xi(x)$ 
(see Ref. \cite{dtok} for discussions of random wall sources in 2- and three-point correlators).

\vspace{.1in}
We fit $C_B^{\beta, \alpha}(t)$ to the form
\begin{eqnarray}
\label{bfit}
C^{\beta, \alpha}_B(t) &=& \sum_{k=0}^{N_B-1} b^\beta_k \, b_k^{\alpha *} 
\,e^{-E^{B,sim}_k \cdot t}  \nl
& & + \sum_{k=0}^{N_B^\prime - 1} b^{\prime \beta}_k \, b_k^{\prime \alpha *} \, (-1)^t \,
 e^{-E^{\prime B, sim}_k \cdot t}
\end{eqnarray}
and 
 $C_D(t; \vec{p})$ to
\begin{eqnarray}
\label{dfit}
C_{D}(t, \vec{p}) &=& \sum_{k=0}^{N_D-1} |d_k|^2 (e^{-E^D_k \cdot t} + e^{-E^D_k
 \cdot ( N_t - t)}) \nl
&+& \sum_{k=0}^{N_D^\prime - 1} |d^\prime_k|^2 (-1)^t 
( e^{-E^{\prime D}_k \cdot t} + e^{-E_k^{\prime D}  \cdot (N_t - t)}).\nl
\end{eqnarray}
The energy $E^{B,sim}_k$ differs from the full energy, $E^B_k$, because the NRQCD action 
has the $b$-quark rest mass removed. For the ground state, the two are related by
\be
E^B_0 \equiv M_B = \frac{1}{2} \left ( \overline{M}_{b \overline{b}}^{exp.} - 
E_{b \overline{b}}^{sim} \right ) + E^{B, sim}_0,
\ee
where $\overline{M}_{b \overline{b}}^{exp.}$ is the spin averaged $\Upsilon$ mass used to tune the $b$-quark mass 
 and suitably adjusted as explained in Sec. II. The values of $E^{sim}_{b \overline{b}}$ can be found in Table I of \cite{bstok}.

\vspace{.1in}
By comparing (\ref{b2pnt}) with (\ref{bfit}) and (\ref{d2pnt}) with (\ref{dfit}), 
and taking the correct relativistic normalisations for the energy eigenstates 
$|E^B_k \rangle$ and $|E^D_k \rangle$ into account, the following relations 
emerge,
\be
\label{bkb}
b_k^{\alpha *} = \frac{\langle E_k^B | \Phi_B^{\alpha \dagger} | 0 \rangle}
{\sqrt{2 a^3 E_k^B}},
\ee
and 
\be
\label{bkd}
d_k = \frac{\langle 0 | \Phi_D | E^D_k \rangle}
{\sqrt{2 a^3 E_k^D}} .
\ee

\begin{figure}
\includegraphics*[scale=0.9, bb=185 550 470 720]{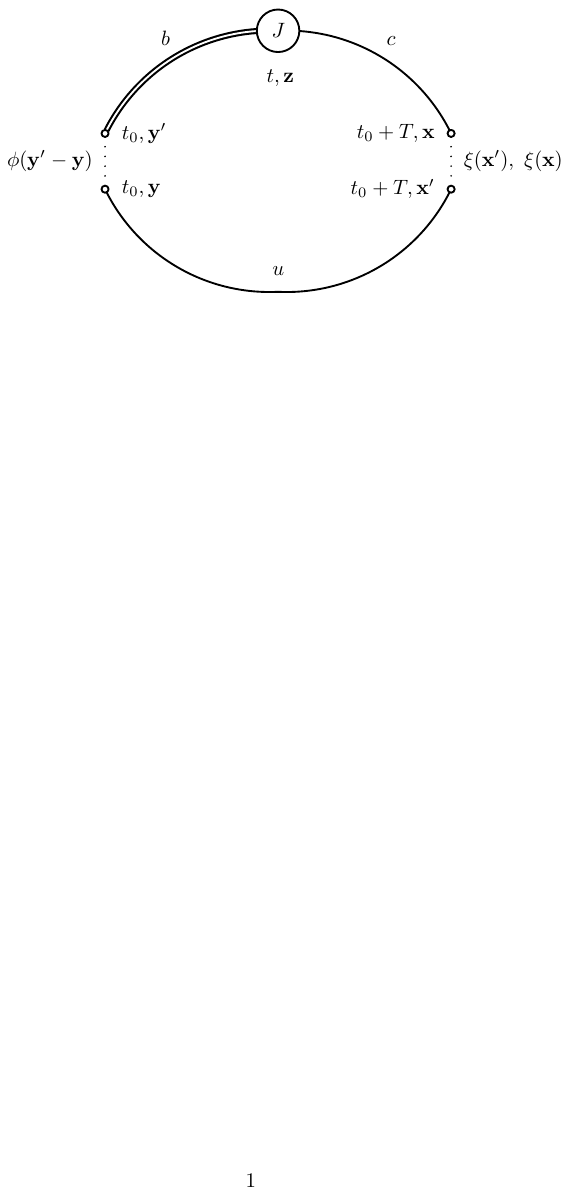}
\caption{
Setup for three-point correlators
 }
\label{dispersion}
\end{figure}

\vspace{.1in}
For the three-point correlator 
 $C^\alpha_J(t,T; \vec{p})$ we use the following fit ansatz
\begin{eqnarray}
\label{thrpntfit}
& &C^\alpha_J (t,T; \vec{p})
 = \sum_{j=0}^{N_B-1} \sum_{k=0}^{N_D-1} A^\alpha_{jk} e^{-E_j^D \cdot t}
 e^{ -E_k^{B,sim} \cdot (T-t)} \nl
& & \quad + \sum_{j=0}^{N_D-1} \sum_{k=0}^{N_B^\prime - 1} B^\alpha_{jk}
 e^{-E_j^{D} \cdot t} e^{ -E_k^{\prime B, sim} \cdot (T-t)} (-1)^{(T-t)} \nl
& & \quad + \sum_{j=0}^{N_D^\prime - 1} \sum_{k=0}^{N_B-1} C^\alpha_{jk}
 e^{-E_j^{\prime D} \cdot t} e^{ -E_k^{B, sim} \cdot (T-t)} (-1)^t \nl
& & \quad + \sum_{j=0}^{N_B^\prime - 1} \sum_{k=0}^{N_D^\prime - 1} D^\alpha_{jk}
 e^{-E_j^{\prime D} \cdot t}
 e^{ -E_k^{\prime B, sim} \cdot (T-t)}  (-1)^{T}. 
\end{eqnarray}
The amplitudes $A^\alpha_{jk}$ etc. depend on the current $J_\mu$ and on the $D$ meson 
momentum $\vec{p}$. Again by comparing (\ref{j3pnt}) and (\ref{thrpntfit}) and 
using (\ref{bkb}) and (\ref{bkd}), one finds
\be
\label{ajk}
A^\alpha_{jk} = d_j \, \frac{\langle E^D_j | J_\mu | E^B_k \rangle } 
{\sqrt{2 a^3 E^D_j} \sqrt{ 2 a^3 E^B_k}} \, b^{\alpha *}_k.
\ee
For $j = k = 0$, $A^\alpha_{00}$ in (\ref{ajk}) gives us the sought after hadronic 
matrix elements $\langle D | J_\mu | B \rangle$,
\be
\label{DJB}
\langle D | J_\mu | B \rangle = \frac{A^\alpha_{00}}{d_0 \, b_0^{\alpha *}} 
\, \sqrt{2 a^3 E^D_0} \, \sqrt{2 a^3 M_B}.
\ee

\begin{figure}
\includegraphics*[width=10.0cm,height=8.0cm,angle=0]{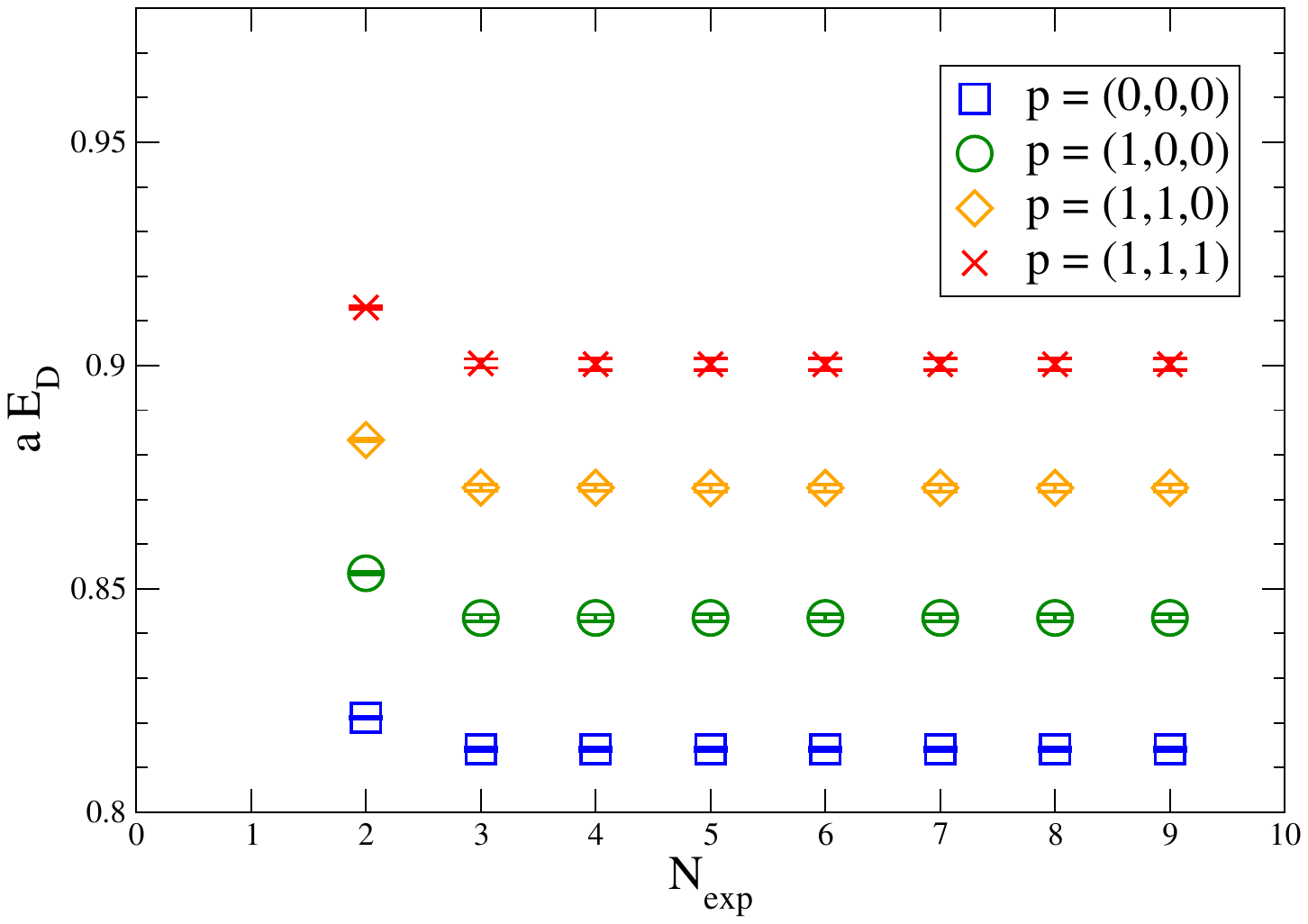}
\caption{
$aE_D$ versus $N_{exp}$ for several momenta and for ensemble F1.
 }
\label{aED}
\end{figure}
\begin{figure}
\includegraphics*[width=10.0cm,height=8.0cm,angle=0]{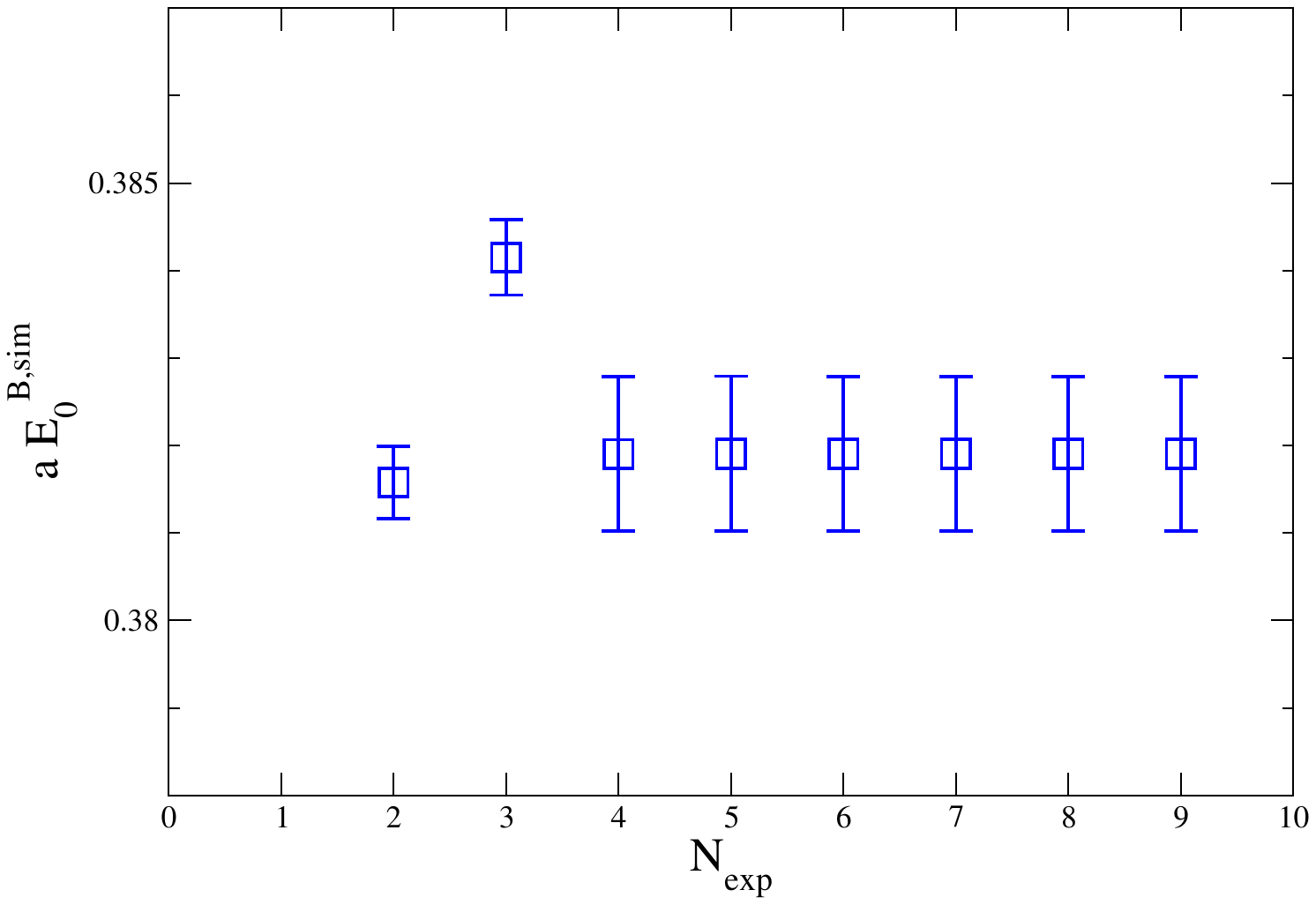}
\caption{
$aE_0^{B,sim}$ versus $N_{exp}$ for ensemble F1.
 }
\label{aMB}
\end{figure}

\begin{figure}
\includegraphics*[width=10.0cm,height=8.0cm,angle=0]{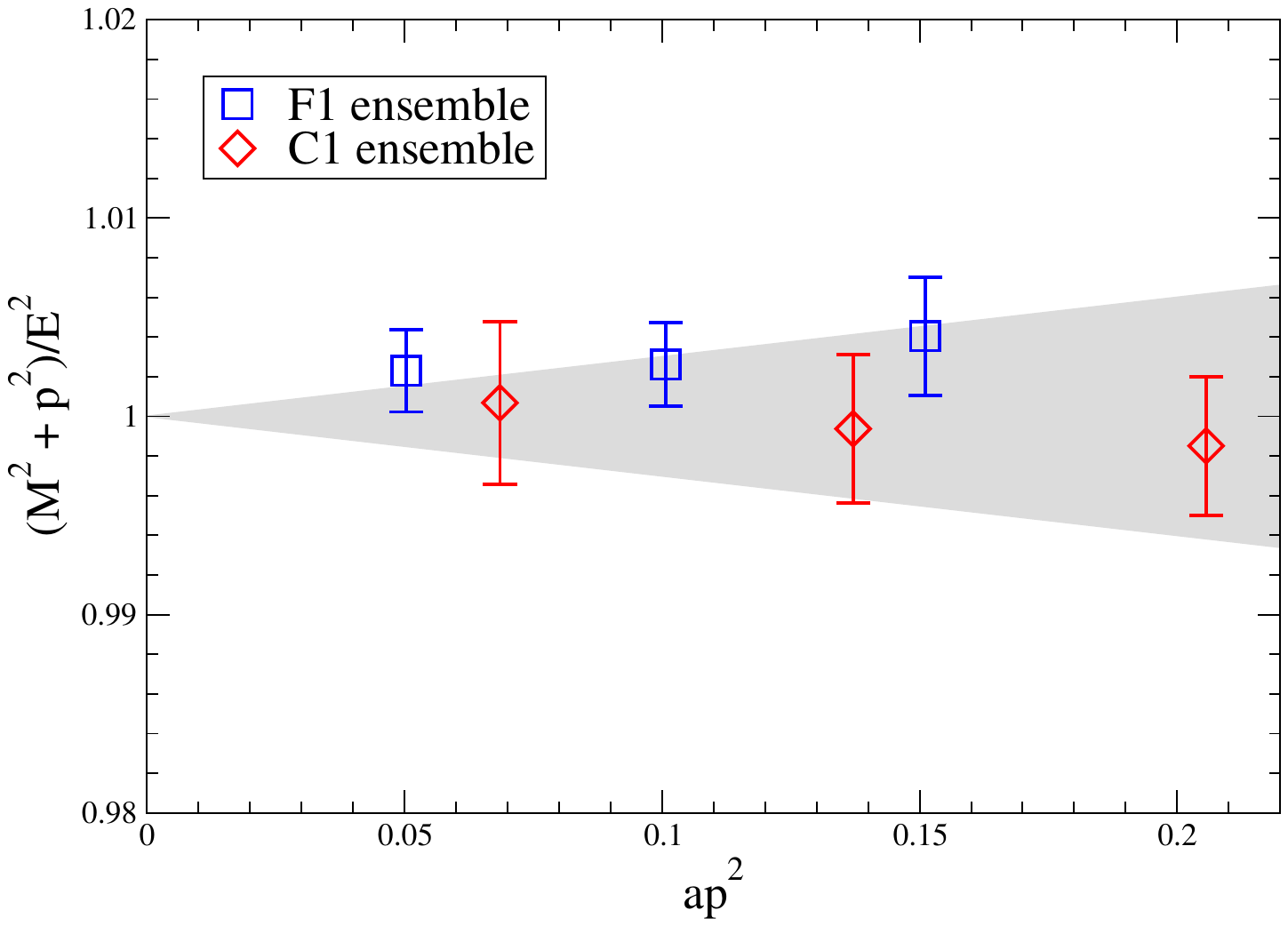}
\caption{
Dispersion relations on ensembles C1 and F1. The shaded region is bounded 
by $1 \; \pm \; c \, \alpha_s \, (ap)^2$ with $c = 0.1$. 
 }
\label{dispersion}
\end{figure}

\begin{figure}
\includegraphics*[width=10.0cm,height=8.0cm,angle=0]{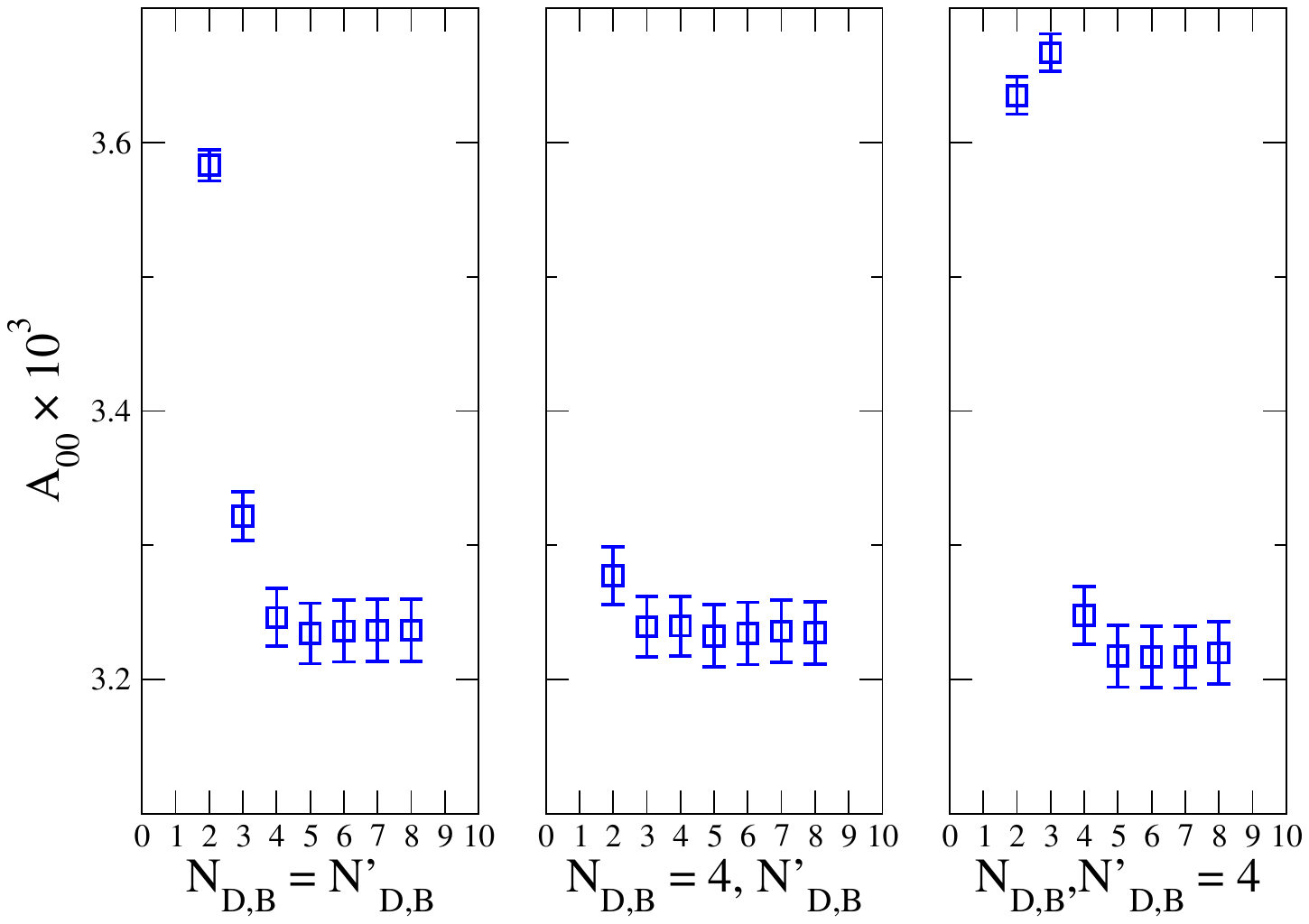}
\caption{
$A_{00}$ for $J_0(\vec{p}=(0,0,0))$ versus $N_{D,B}$ and $N^\prime_{D,B}$ for F1. The two plots on the right are at fixed $N_{D,B} = 4$ or $N^\prime_{D,B} = 4$.
 }
\label{a00.1}
\end{figure}

\begin{figure}
\includegraphics*[width=10.0cm,height=8.0cm,angle=0]{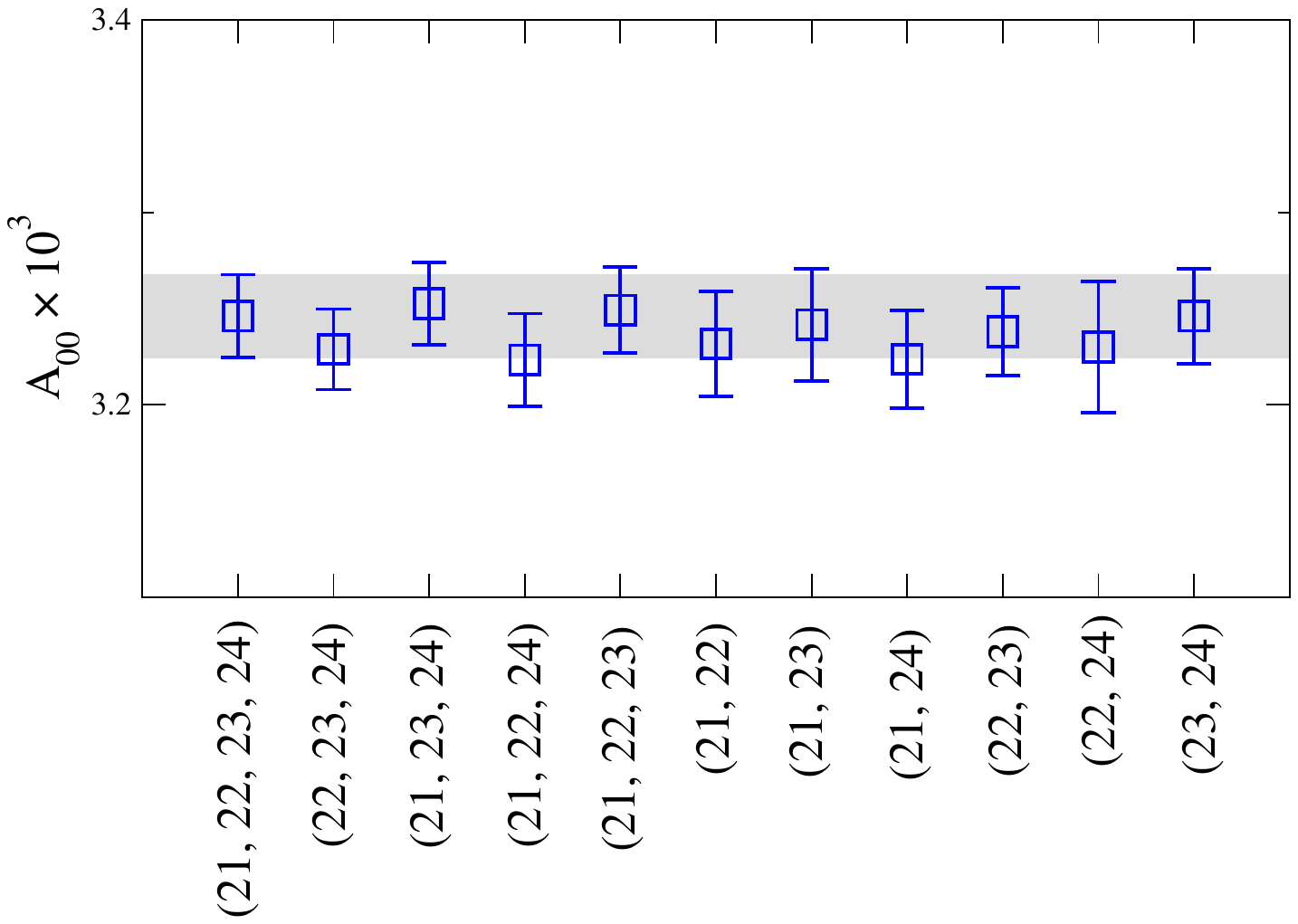}
\caption{
$A_{00}$ for $J_0(\vec{p}=(0,0,0))$ versus different $T$ combinations for F1. We take the result with $T=(21,22,23,24)$.}
\label{a00.2}
\end{figure}

\vspace{.1in}
Our fitting strategies based on Bayesian methods have been developed and refined 
in a number of calculations~\cite{fit1,fit2}.  We follow closely the approach used in our recent 
$B_s \rightarrow K l \nu$ studies \cite{bstok}. 
Figs. 2 and 3 show fit results for the groundstate $D$ meson energies $a E_D$ 
and for  $a E_0^{B,sim}$, respectively, versus the number of exponentials 
in the fit $N_{exp}$ (we set $N_{exp} = N_{D,B} = N^\prime_{D,B}$).  
Fits stabilize after $N_{exp} = 4$.  In Fig.~4 we check the ratio 
$(M^2 + p^2) / E^2$ for the $D$ meson on one coarse (C1) and one fine (F1) 
ensemble. The shaded area is bounded by
 $1 \; \pm \; c \: \alpha_s (ap)^2$, where the free parameter
 $c$ has been set to $0.1$.
  One sees that the relativistic dispersion relation
 holds within errors to better than 0.5\%.

\vspace{.1in} 
For fixed $D$ momentum, the combination on the rhs of Eq.~(24) is obtained from a simultaneous fit to a $2 \times 2$ matrix of $B$-correlators, one $D$ correlator and numerous three-point correlators.  The number of three-point correlators in the fits varies from six to sixteen as we use either three or four values of $T$, two smearings $\alpha$, and either one current at zero momentum ($J_0$) or two currents at non-zero momenta ($J_0$ and $J_i$).
We call this type of fit an ``individual fit''. 
Fig.~5 shows results for $A^\alpha_{00}$ for $J_0(\vec{p}=(0,0,0))$ versus the number of exponentials. 
And Fig.~6 shows how results depend on choices for 
different  $T$  combinations. Individual fits 
give stable and consistent results under such variations.

\begin{figure}
\includegraphics*[width=10.0cm,height=8.0cm,angle=0]{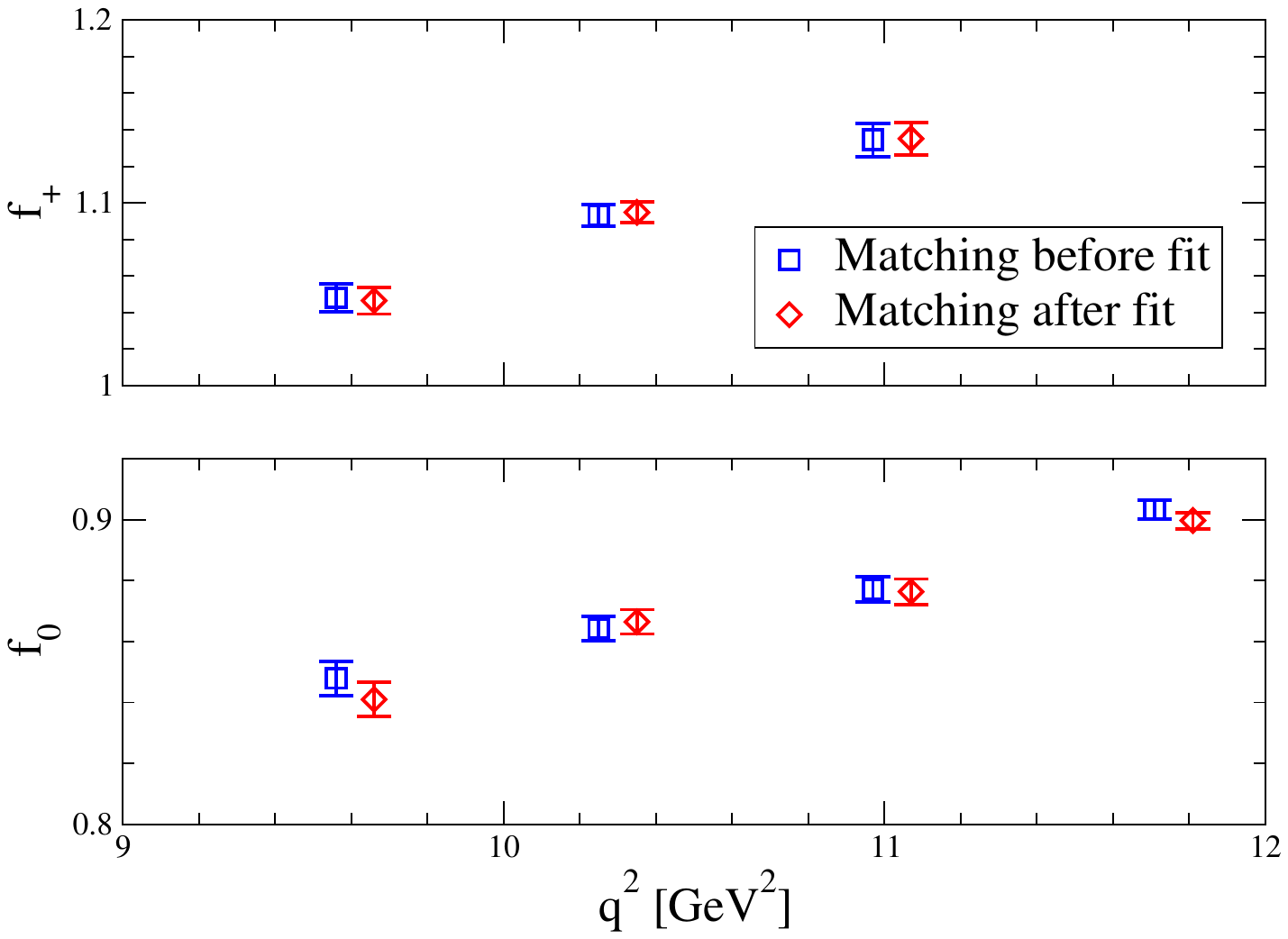}
\caption{
Comparison between form factor fit results when  current matching corrections are 
undertaken before or after the fits (for ensemble F1). }
\label{matching}
\end{figure}

\vspace{.1in}
We fit data after the current matching described in Sec. II.  
Specifically, we obtain simulation data for $J_\mu^{(0)}$ and $J_\mu^{(1)}$ of Eqs.~(8) and (9), reconstruct the full expression on the RHS of Eq~(10), and fit the resulting data.  Alternatively, we can perform separate fits to $J_\mu^{(0)}$ and $J_\mu^{(1)}$ and then combine the results according to Eq.~(10).  We have compared the two approaches and find good agreement, as shown in Fig.~\ref{matching}.

\begin{figure}
\includegraphics*[width=10.0cm,height=8.0cm,angle=0]{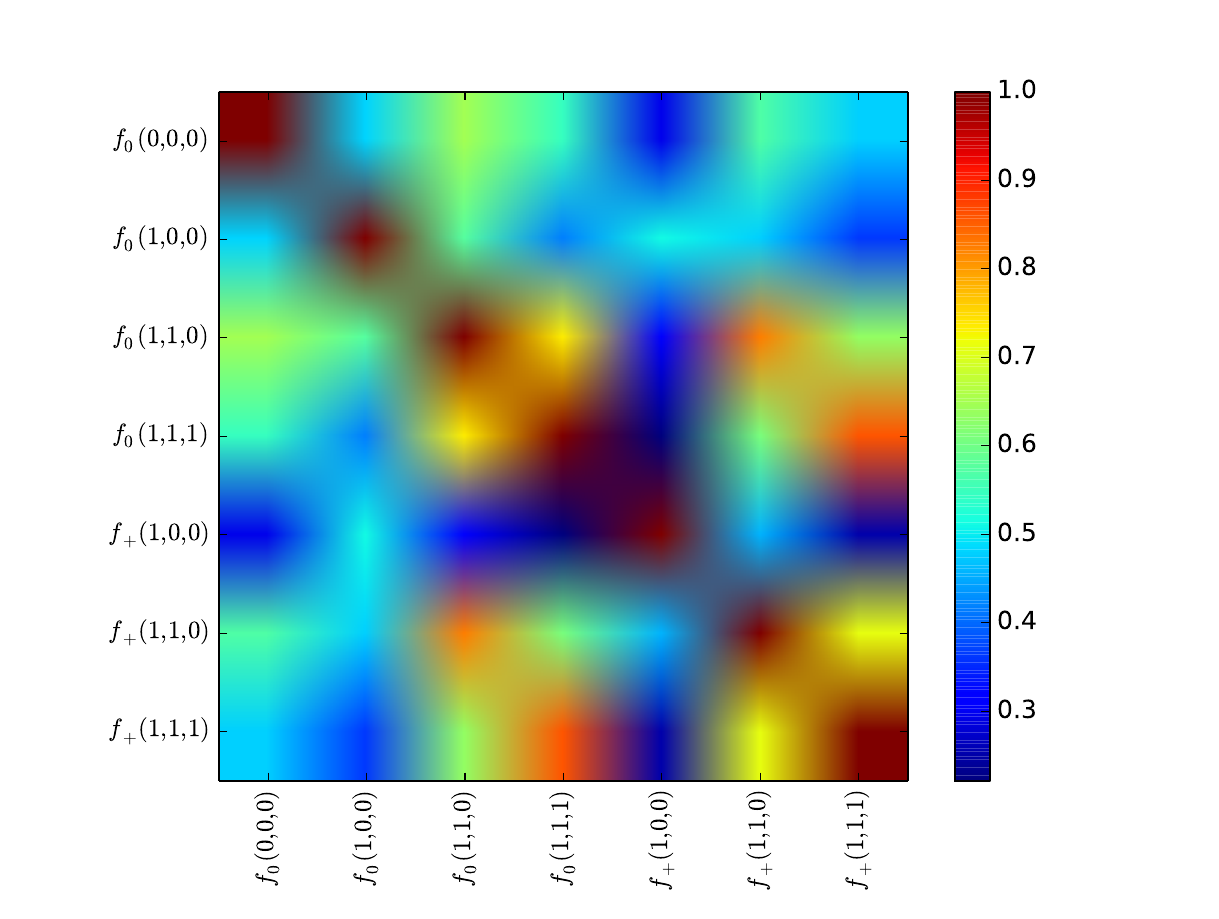}
\caption{
Correlations between different momenta from the master fit for ensemble F1. }
\label{correl}
\end{figure}

\vspace{.1in}
In order to get correlations of form factors at different $q^2$'s, one needs to do a simultaneous fit with all different $D$ momenta.
Each individual fit alone involves 10 to 20 
correlators (depending on the combination of $T$ values) and a simultaneous fit
 requires a four times larger set of  correlators. 
 We find that the simultaneous fits lead to unreliable results,
 indicating that they are too complicated given the accuracy of our data.
 Our fitting routines, however, allow for an alternate way to keep track
 of correlations between form factors at different $q^2$'s.
 One can do a sequence of individual fits, one after the other,  all within a single script
 and always employing the full covariance matrix for all the data
 (all $D$ meson momenta). We call such fits ``master fits''. These master
 fits are easier than straight simultaneous fits,
 but still highly non-trivial and  time consuming.
It was possible to get good master fits 
consistent with individual fit results; however, these were less stable 
with respect to changes in $N_{exp}$ and $T$ combinations.  Hence 
for our final fit results we use central values and errors from 
individual fits, and use the good master fits just to extract the necessary 
correlations.  
Fig.~\ref{correl} shows an example of correlations obtained from a master fit 
to all the data from ensemble F1. 
Table III summarizes form factor results for each 
ensemble and $D$ momentum, and Table IV shows the fit results of $B$ and $D$ meson ground state energies.

\begin{table}
\caption{
Fit results for $f_0(\vec{p})$ and $f_+(\vec{p})$.
}
\begin{ruledtabular}
\begin{tabular}{ccccc}
Set &  $f_0(0,0,0)$ & $f_0(1,0,0)$   &  $f_0(1,1,0)$&
 $f_0(1,1,1)$  \\
\hline
C1  & 0.8810(56) &  0.8743(43)  & 0.8608(38)  & 0.8534(42)   \\
C2  & 0.8809(31) & 0.8716(54)  &  0.8617(44)  & 0.8503(50)  \\
C3  & 0.8872(23)  &  0.8685(32) & 0.8592(29)  &  0.8473(38) \\
\hline
F1  & 0.9034(31) & 0.8771(42) &  0.8643(41) &  0.8479(56) \\
F2  & 0.9051(23) & 0.8895(36) & 0.8702(29)  & 0.8504(34)  \\
\end{tabular}
\begin{tabular}{ccccc}
Set  & $f_+(1,0,0)$   &  $f_+(1,1,0)$&
 $f_+(1,1,1)$  \\
\hline
C1  & 1.135(12) &  1.1125(57)  &    1.0837(61)   \\
C2  & 1.110(12) & 1.0809(70)  &    1.0479(64)   \\
C3  & 1.1282(71) &  1.0937(40)  &  1.0569(50)    \\
\hline
F1  & 1.1344(91) & 1.0931(59) &  1.0480(74)    \\
F2  & 1.1461(72) & 1.0963(39) & 1.0577(45)     \\
\end{tabular}
\end{ruledtabular}
\end{table}

\begin{table}
\caption{
Fit results for $aE_D$ with each momentum and $aE_0^{B,sim}$.
}
\begin{ruledtabular}
\begin{tabular}{cccccc}
Set   & $aE_D(0,0,0)$   & $aE_D(1,0,0)$ & $aE_D(1,1,0)$  & $aE_D(1,1,1)$ \\
\hline
C1   & 1.1388(15)  & 1.1681(19) & 1.1979(17) & 1.2267(16)  \\
C2  & 1.1577(14) & 1.1959(29) & 1.2365(28)  & 1.2758(34) \\
C3   & 1.16355(69) & 1.2046(11) & 1.2442(12) & 1.2824(18) \\
\hline
F1  &  0.81409(42) & 0.84349(77)  & 0.87262(83)  &  0.9003(13) \\
F2  & 0.81999(37) & 0.85051(69)  &  0.87905(66) & 0.90682(86) \\
\end{tabular}

\begin{tabular}{cc}
Set & $aE_0^{B,sim}$   \\
\hline
C1  & 0.4964(13)   \\
C2  & 0.5089(14)  \\
C3  & 0.51376(95)  \\
\hline
F1  & 0.38190(89)  \\
F2  & 0.38726(73) \\
\end{tabular}
\end{ruledtabular}
\end{table}

\section{ Chiral, Continuum and Kinematic Extrapolation }

In this section we describe how we extrapolate the form factors of 
Table III to the continuum and chiral limits, and how 
one can get information on the form factors for the entire physical 
kinematic range. In the continuum physical theory, form factors are 
functions of a single kinematic variable which can be taken 
to be $q^2$, $E_D$, $(w - 1) \equiv (q^2_{max} - q^2)/(2 M_B M_D)$ or 
the {\it z-variable} defined in terms of $q^2$ as,
\be
\label{zvar}
z(q^2) = \frac{\sqrt{t_+ - q^2} - \sqrt{t_+ - t_0}}  
{\sqrt{t_+ - q^2} + \sqrt{t_+ - t_0}}.
\ee
Here $t_+ = (M_B + M_D)^2$ and 
$t_0$ is a free parameter which we set to $t_0 = q^2_{max} = (M_B - M_D)^2
 \sim 11.66$ GeV$^2$. 
A popular expansion in terms of $z$ is the Bourrely-Caprini-Lellouch (BCL) parametrization \cite{bcl}, which is
 given as
\be
\label{bcl1}
f_+(q^2) = \frac{1}{P_+} \sum_{k=0}^{K-1} a_k^{(+)}[ z(q^2)^k - (-1)^{k-K}\frac{k}{K}z(q^2)^K ]
\ee
and,
\be
\label{bcl2}
f_0(q^2) = \frac{1}{P_0}\sum_{k=0}^{K-1} a_k^{(0)} z(q^2)^k.
\ee
Here $P_{+,0}$ are the Blaschke factors that take into account the effects of 
expected poles above the physical region but below the two body threshold $t_+$, 
i.e. in the region $(M_B - M_D)^2 < q^2 < (M_B + M_D)^2$,
\be
P_{+,0}(q^2) = \left ( 1 - \frac{q^2}{M_{+,0}^2} \right ).
\ee
For $f_+$ we take the $B_c^*$ vector meson mass which has been calculated 
in Ref.~\cite{bcstar}, $M_+ = M_{B_c^*} = 6.330(9)$GeV.  For the scalar form factor $f_0$, there 
is little information theoretically or experimentally
on a $0^+$ bottom-charm meson. We take $M_0$ to be slightly heavier than $M_+$ 
with large errors. 
 We find that our 
fit results are very insensitive to our choice of $M_0$. Even omitting the 
Blaschke factor completely for $f_0(q^2)$ leads to results consistent with
 keeping it in (see test number 16 below). 
The poles in the $B \rightarrow D l \nu$ form factors 
 are located  far above the physical $q^2$ region, for example $q^2_{max} = (M_B-M_D)^2 \sim 11.6 \; \text{GeV}^2$ while $M_{B_c^*}^2 \sim 40 \; \text{GeV}^2$.  
This implies that the form factors have very small curvatures, and in fact it is very difficult to quantify the curvature for $f_0$ from our lattice data. 

\vspace{.1in}
Once the contributions from simple poles have been isolated, the 
power series in (\ref{bcl1}) and (\ref{bcl2}) correspond to 
smooth functions of $z$.  One reason for prefering a power series 
in $z$, as opposed to one in $q^2$ or $E_D$, or even $(w -1 )$ is
   that $|z|$ remains very 
small throughout the physical kinematic region. For $B \rightarrow D$ 
semileptonic decays and our choice for 
$t_0$, one has $0.0 \leq z < 0.064$.  
This means that one can go to arbitrary high powers in $z^k$ if 
necessary (in practice, with our current simulation data, going up 
to $z^3$ will suffice).

\vspace{.1in}
The form factors of Table III are not yet in the physical limit.  Nevertheless, 
for fixed lattice spacing and pion mass, one can again write form factors 
in terms of a Blaschke factor mutliplying a power series in $z$.  The advantages 
of the z-expansion relative to an expansion in, for instance, powers of $E_D$ 
still hold away from the physical limit. What is different, however, is 
that expansion coefficients must now depend on the lattice spacing ``$a$'' 
and on ``$m_\pi$'' (or the light quark mass),
\be
\label{zexpan1}
a_k^{(0,+)} \rightarrow \tilde{a}_k^{(0,+)} \times D_k^{(0,+)}(m_l, m_l^{sea}, a),
\ee
with
\be
a_k^{(0,+)} = \tilde{a}_k^{(0,+)} \times D_k^{(0,+)}(m_l(phys.), m_l^{sea}(phys.), 
a = 0).
\ee
This is the modified z-expansion first introduced in Ref. ~\cite{dtok, dtopi} 
for $D$ meson semileptonic decays, and which has subsequently also 
 been employed successfully 
 in $B$ and $B_s$ meson heavy-to-light decays \cite{btok1,btok2,bstok}.
The $D_k$ in (\ref{zexpan1})  contains all lattice artifacts and chiral logs.
Specifically, we have,
\begin{eqnarray}
\label{dk}
D_k &=& 1 + c_1^k x_\pi + c_2^k (\frac{1}{2} \delta x_\pi 
 + \delta x_K) + 
c_3^k x_\pi {\rm log}(x_\pi)  \nonumber \\
&+& d_1^k (am_c)^2 + d_2^k (am_c)^4  \nonumber \\
&+& e_1^k (aE_D/\pi)^2 + e_2^k (aE_D/\pi)^4, 
\end{eqnarray}
where
\begin{eqnarray}
x_{\pi,K,\eta} &=& \frac{M_{\pi,K,\eta}^2}{(4\pi f_\pi)^2}, \\
\delta x_{\pi,K} &=& \frac{(M_{\pi,K}^{asqtad})^2 - (M_{\pi,K}^{HISQ})^2}
{(4\pi f_\pi)^2}.
\end{eqnarray}
The $c^k_j$, with $j = 1,2,3$, and $d^k_i$ and $e^k_i$, with $i = 1,2$, 
are fit parameters (we have omitted the $f_{+,0}$ 
label for simplicity) 
in  addition to the $\tilde{a}_k^{0,+}$.  In Appendix B we discuss what happens 
when the simple chiral log term in (\ref{dk}) is replaced by 
expressions from Hard Pion Chiral Perturbation Theory (HPChPT) \cite{hpchpt} 
(see also test number 10 below).
We list the priors and prior widths used in 
the chiral/continuum/kinematic extrapolation in Appendix C.

\vspace{.1in}
We find it useful to make one more modification of 
the z-parametrization of lattice form factors.
 In order to accommodate the uncertainty coming from 
the truncation of the current matchings at ${\cal O}(\alpha_s, \Lambda_{QCD}/M, 
\alpha_s/(aM))$,
 we introduce new fit parameters, $m_\|$ and $m_\bot$, 
with central value zero and width $\delta m_{\|,\bot}$,
 \be
\label{matcherror}
 f_\|, f_\bot \rightarrow (1+m_\|)f_\|, (1+m_\bot)f_\bot.
 \ee
The prior widths $\delta m_\|$ and $\delta m_\bot$ correspond
 to our best estimates for higher order 
matching errors for $V_0$ and $V_k$ respectively.
 With the modification of (\ref{matcherror}), our extrapolation 
results coming from the modified z-expansion fit will then 
 include the matching truncation errors automatically.  
To get an estimate of higher order matching uncertainties and fix 
$\delta m_{\|,\bot}$, we have looked 
at the size of the known first order matching corrections.  In other words  
we have gone through the correlator fits of the previous section 
once using the fully corrected expression on the rhs of (\ref{jmu}) and then 
a second time using just the lowest order $\langle J^{(0)}_\mu \rangle$.  
We find that the first order matching contributions have only a $\sim 2$\% 
effect on fine and a $\sim 4$\% effect on coarse lattices, 
 significantly smaller than a naive $1 \times {\cal O}
(\alpha) \approx 25 - 30$\% estimate.  In this work we take the higher order 
uncertainties to be the same as the average of the full first order corrections
  on fine and coarse lattices, that is 
we set the prior central values and widths 
of the fit parameters $m_{\|,\bot}$ to be $0.0 \pm 0.03$.  
We have checked that using $0.0 \pm 0.02$ or $0.0 \pm 0.04$ everywhere,  or 
$0.0 \pm 0.02$ for fine and $0.0 \pm 0.04$ for coarse lattices 
has minimal effect (see tests number 13, 14, and 15 below).
After the modified z-expansion fits and extrapolation to the 
physical limit, these matching uncertainties for $f_\|$ and $f_\bot$ 
will translate into matching errors for $f_+$ and $f_0$ with correlations 
between the two form factors taken into account.

\begin{figure}
\includegraphics*[width=10.0cm,height=8.0cm,angle=0]{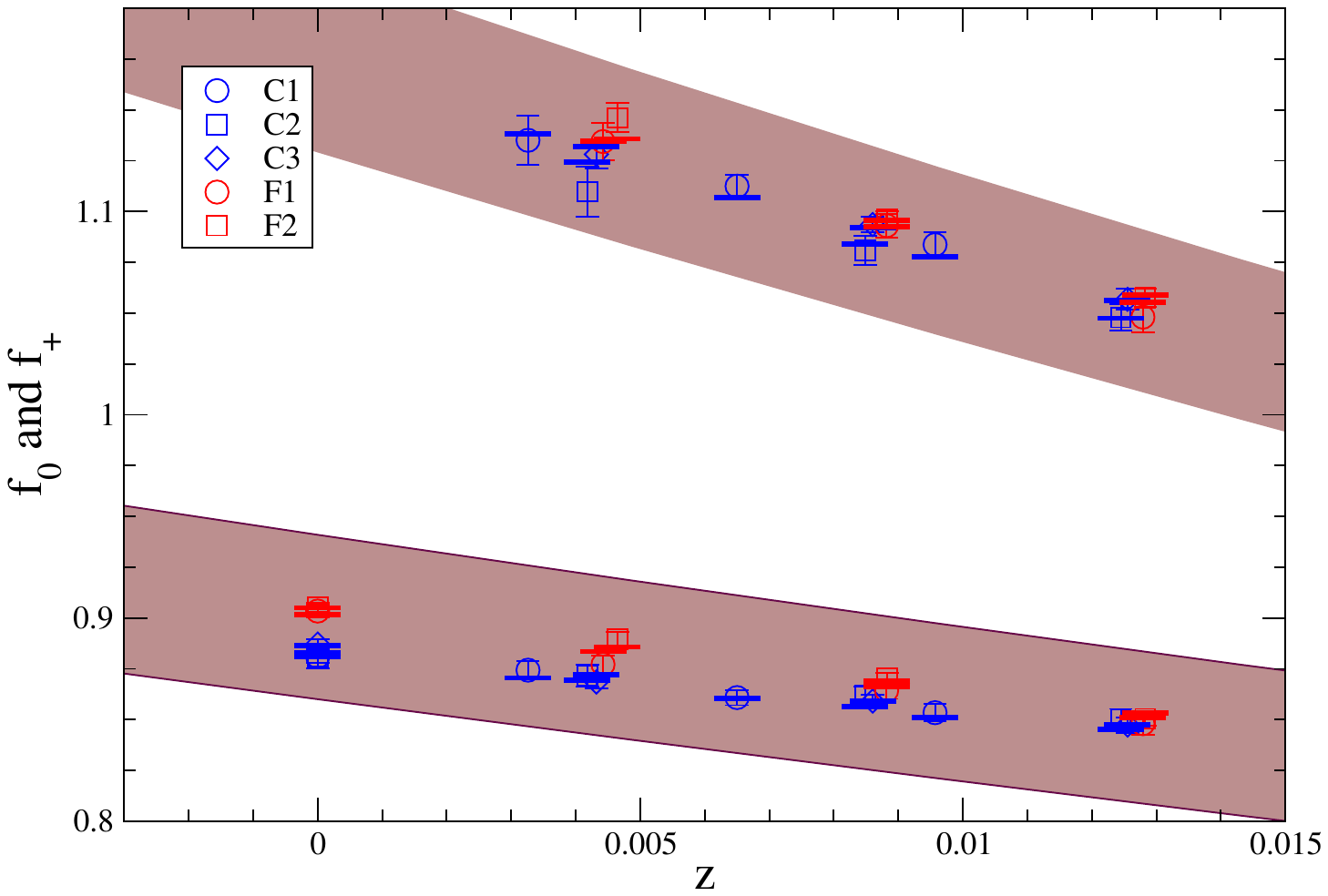}
\includegraphics*[width=10.0cm,height=8.0cm,angle=0]{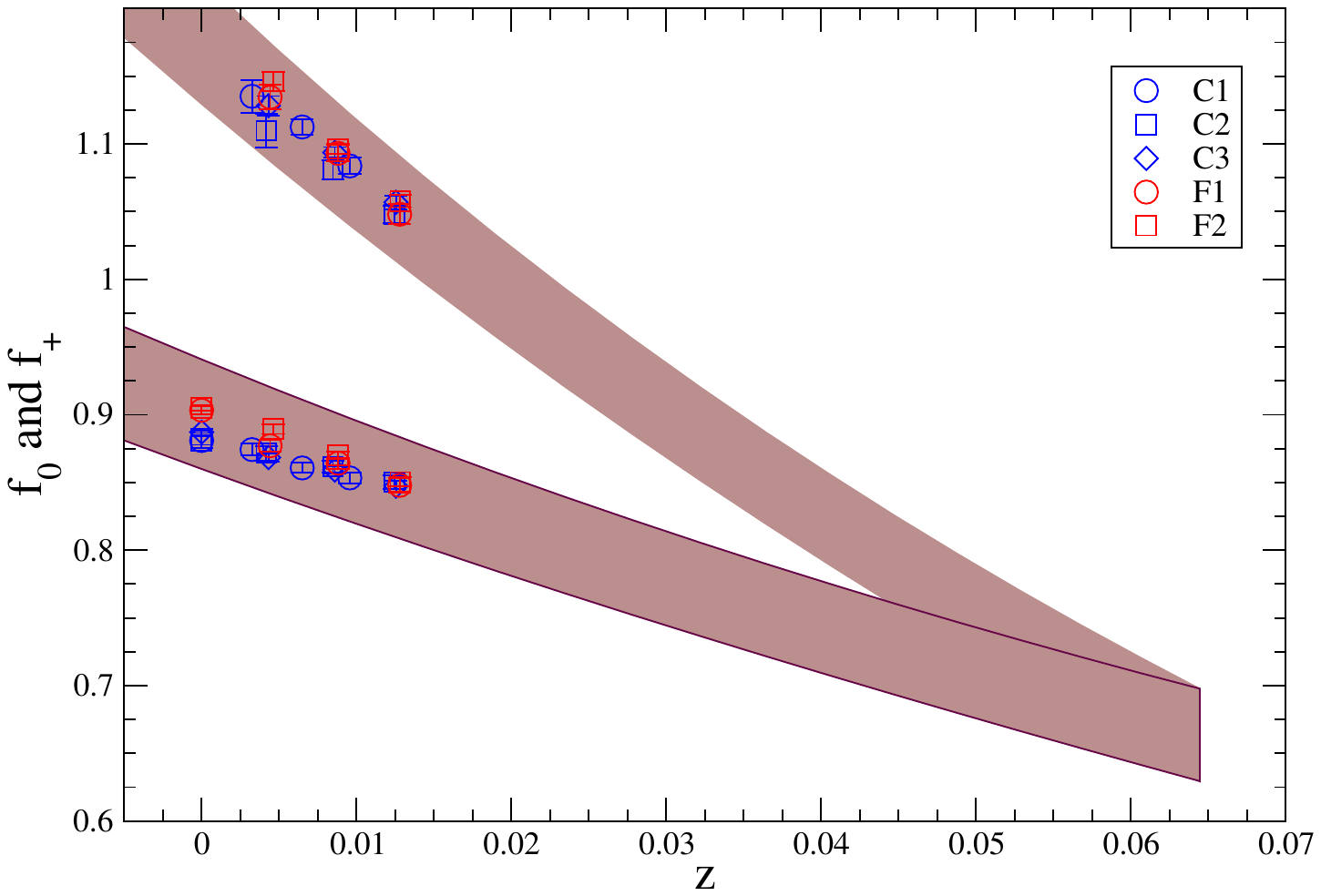}
\caption{
The standard fit results with the continuum extrapolated bands. The short horizontal bars on the upper plot show the fit results at non-zero lattice spacings. }
\label{f0fpz}
\end{figure}

\vspace{.1in}
In Fig.~\ref{f0fpz} we show our fit results for $f_+$ and $f_0$ plotted versus $z$. We plot both the simulation data and  
the extrapolated physical band. These are results of what we call our 
``standard extrapolation'' which uses the fit ansatz discussed above and 
a z-expansion that includes terms through ${\cal O}(z^3)$. 
We have carried out further tests of the standard extrapolation by 
modifying the fit ansatz in the following ways:
\begin{enumerate}
\item
stop at ${\cal O}(z^2)$ in the z-expansion;

\item
stop at ${\cal O}(z^4)$ in the z-expansion;

\item
add light quark mass dependence to $d_1^k$ (see Eq.~(30) of  \cite{bstok});

\item
add bottom quark mass dependence to $d_1^k$ (see Eq.~(30) of  \cite{bstok});

\item
omit $(am_c)^4$ term;

\item
add $(am_c)^6$ term;

\item
omit $(a E_D/\pi)^4$ term;

\item
add $(a E_D/\pi)^6$ term;

\item
omit $x {\rm log}(x) $ term;

\item
use chiral logs from HPChPT (see Appendix B);

\item
add $x_\pi^2$ term;
 
\item
omit all $x_i$ and $x {\rm log}(x)$ terms;

\item 
use 2\% uncertainty for higher order matching contributions;

\item 
use 4\% uncertainty for higher order matching contributions;

\item 
use 2\% uncertainty on fine and 4\% uncertainty 
on coarse lattices for higher order matching contributions;

\item
remove Blaschke factor from $f_0$ and $f_+$.

\end{enumerate}

\begin{figure}
\includegraphics*[width=10.0cm,height=8.0cm,angle=0]{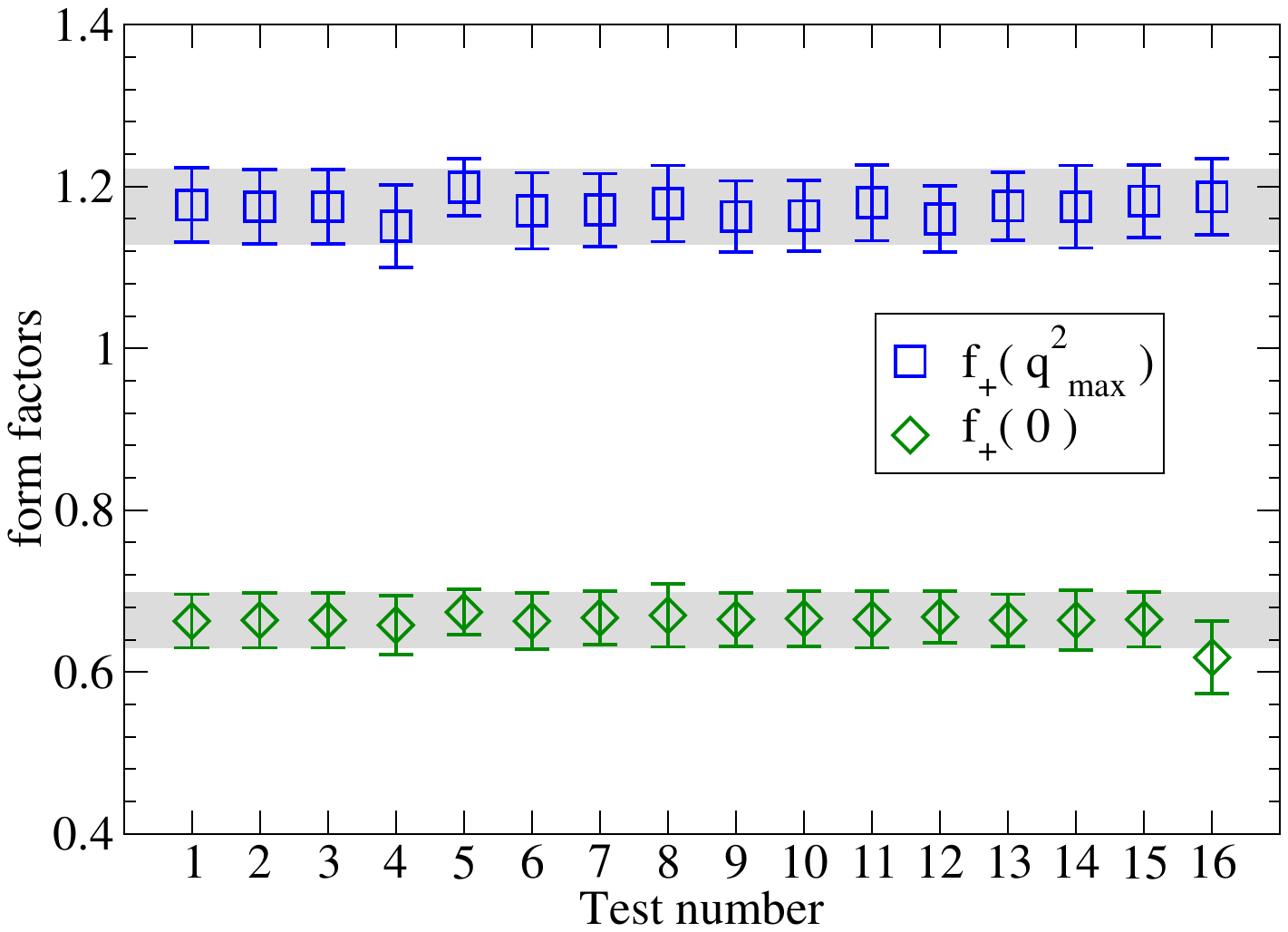}
\caption{
Test results for $f_+(0)$ and $f_+(q^2_{max})$ under modifications 
of the ``standard extrapolation'' fit ansatz. The 
shaded horizontal bands are the standard extrapolation results.  The x-axis 
labels the modifications 1 - 16 listed in the text.
}
\label{tests}
\end{figure}

In Fig.~\ref{tests} we show how results for $f_+(q^2=0) = f_0(0)$ and $f_+(q^2_{max})$
 are affected by 
these modifications.  One sees that our extrapolations are very 
stable.

\section{ Form Factor Results}

\begin{figure}
\includegraphics*[width=10.0cm,height=8.0cm,angle=0]{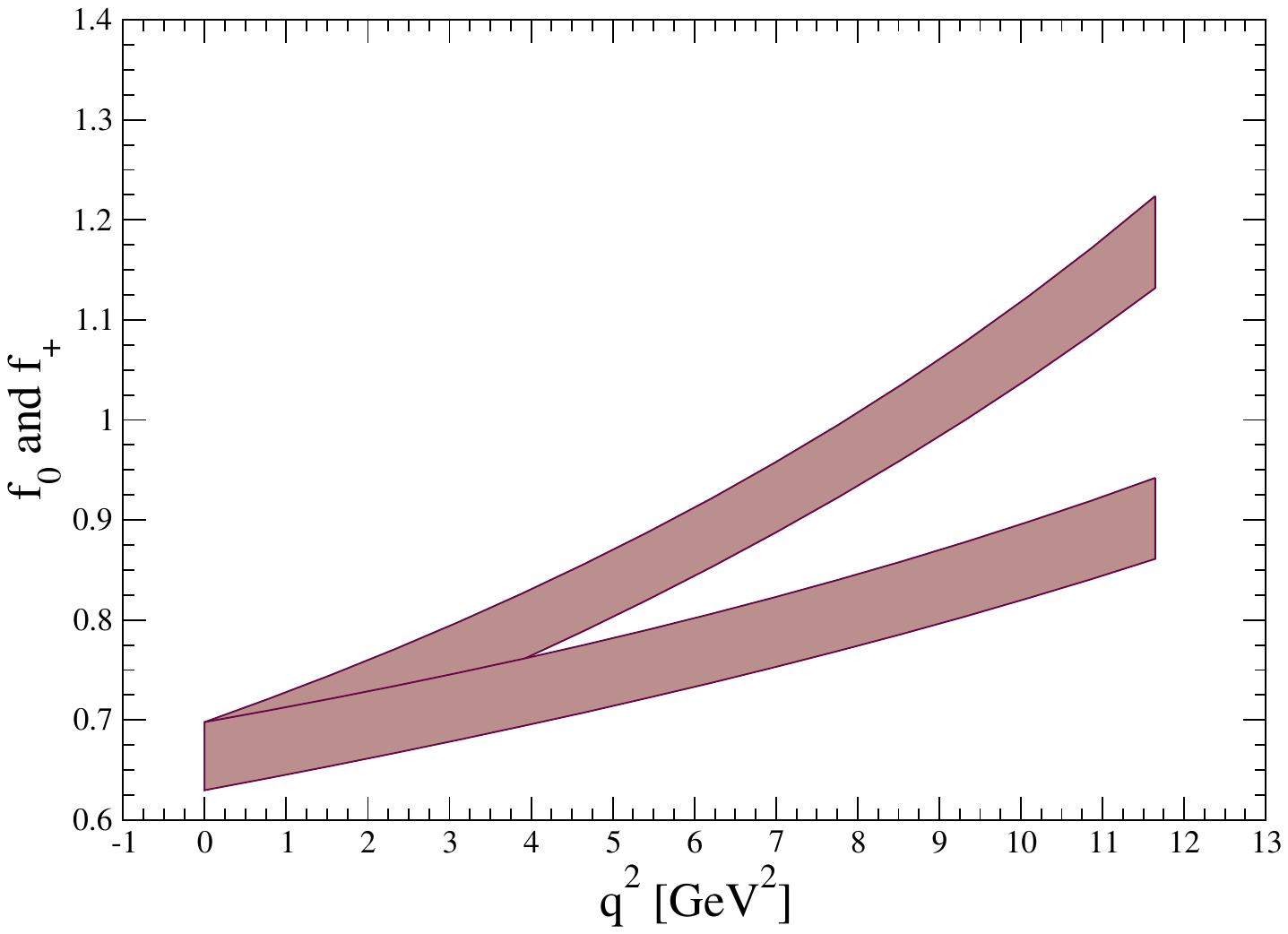}
\caption{
Continuum and chiral extrapolated $f_0$ (lower band) and $f_+$ (upper band).}
\label{f0fp}
\end{figure}

\begin{figure}
\includegraphics*[width=10.0cm,height=8.0cm,angle=0]{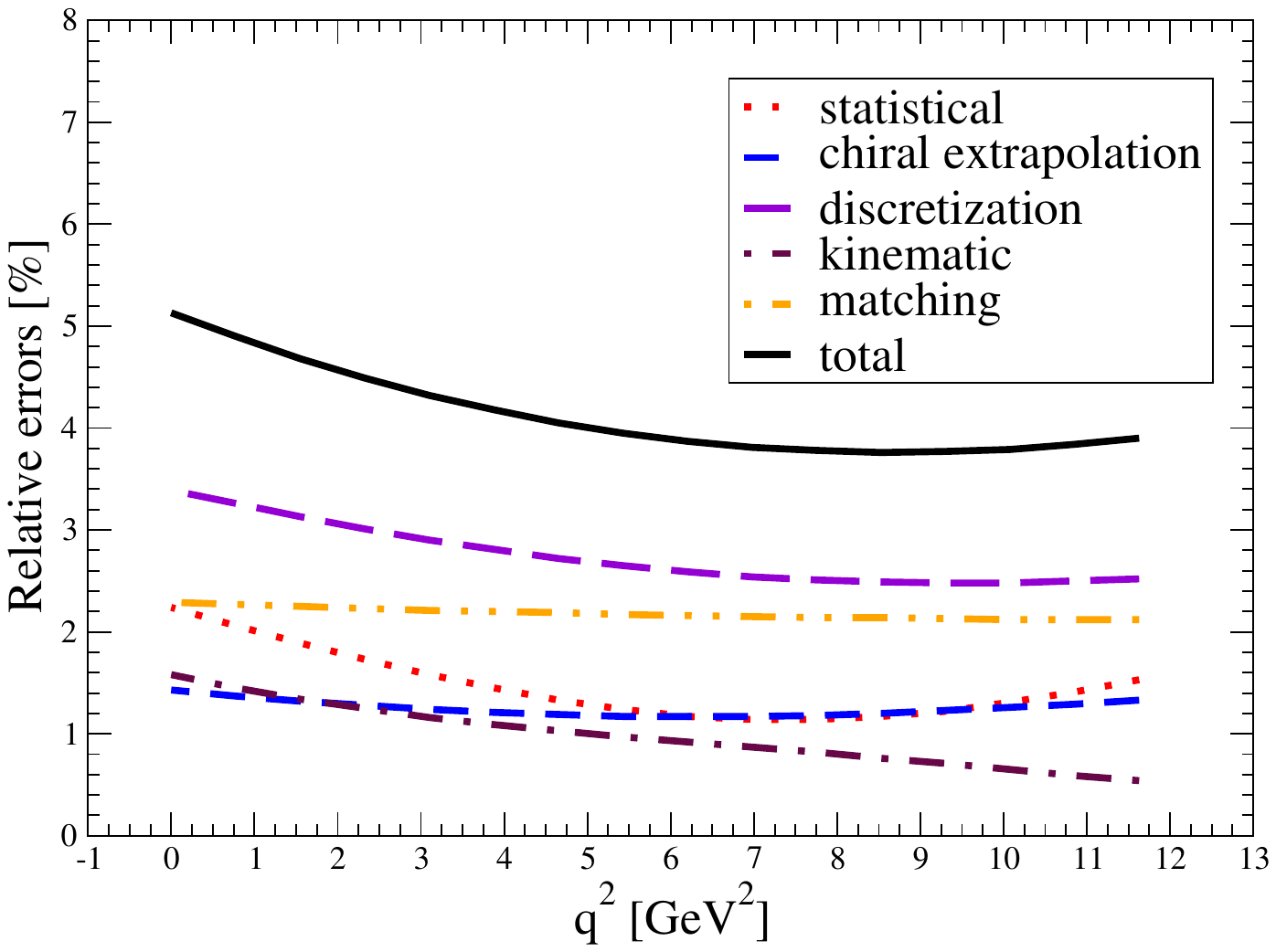}
\includegraphics*[width=10.0cm,height=8.0cm,angle=0]{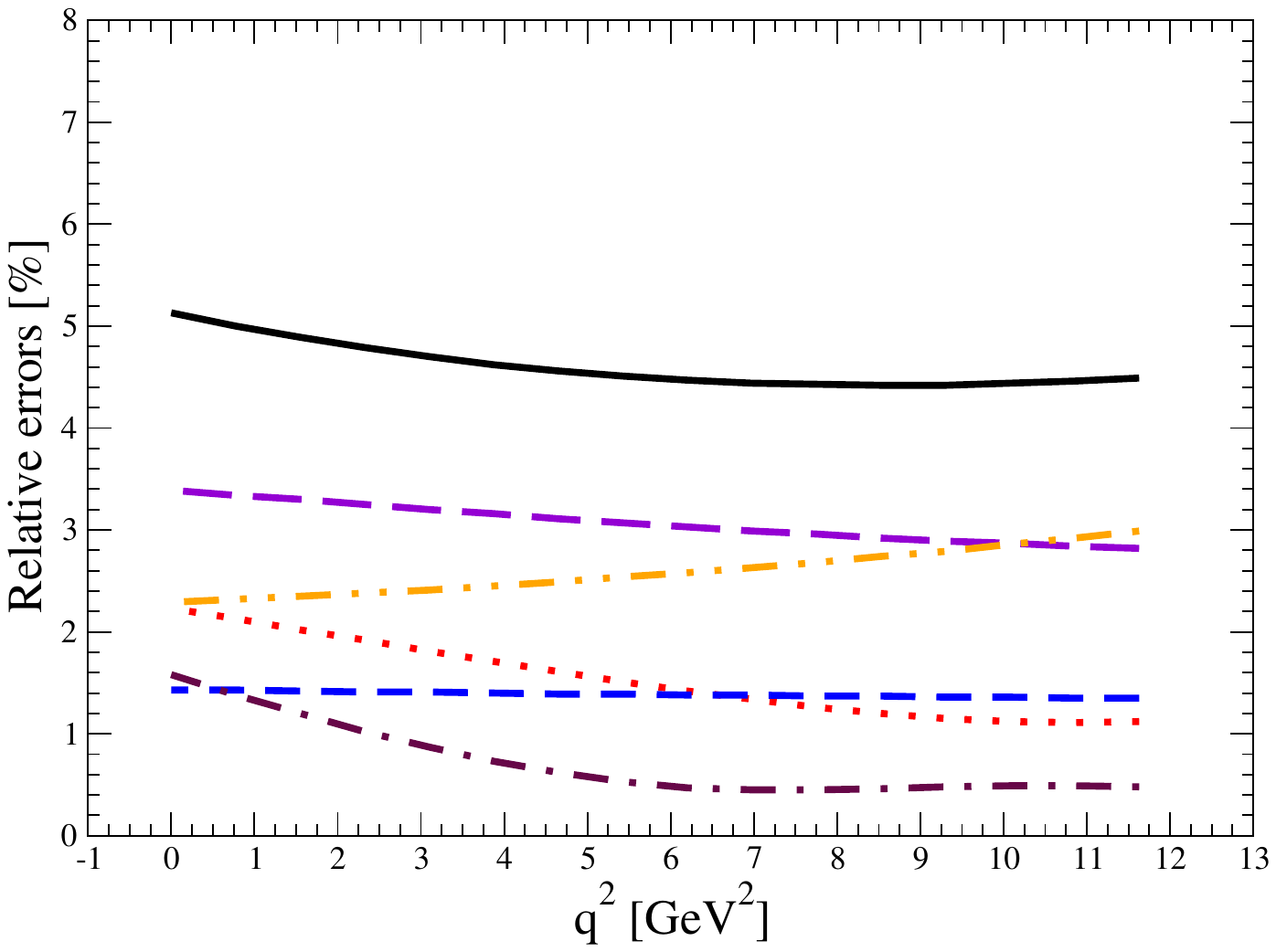}
\caption{
Relative error components of  $f_0$ (lower plot) and $f_+$ (upper plot) for physical $q^2$ region.}
\label{error}
\end{figure}



Our final results for the form factors in the physical limit versus $q^2$ are shown 
in Fig.~\ref{f0fp}. Error plots for $f_+(q^2)$ and $f_0(q^2)$ are given 
in Fig.~\ref{error}.  We isolate the errors coming from different sources and 
also give the total error as a function of $q^2$.
The individual errors in Fig.~\ref{error} correspond to the following:
 \renewcommand{\labelitemi}{$\bullet$}
\begin{itemize}
\item {\bf statistical} \\
The statistical error includes the three and two-point correlator fit errors and the scale errors ($r_1$ and $r_1/a$).
These are lattice simulation errors, and we have lattice data in the large $q^2$ region, from about $9.5 ~\text{GeV}^2$ to $12 ~\text{GeV}^2$.  Fig.~\ref{error} shows the propagation of such errors to the continuum limit and after extrapolation to the full $q^2$ range. 
\item {\bf chiral extrapolation}\\
These are the valence and sea quark mass extrapolation errors
 including effects of chiral logs. They
 come from the fit parameters $c^k_1$, $c^k_2$ and $c^k_3$ in Eq.~(\ref{dk}).
\item{\bf discretization}\\
Discretization errors come from the  $(am_c)^n$ and $(aE_D)^n$ terms and they 
constitute the dominant errors in our calculation.  
\item{\bf kinematic}\\
 These come from the z-expansion coefficients $\tilde{a}^{(0,+)}_k$ and 
the pole locations.  As one would expect, the error increases as $q^2$ decreases. 
\item{\bf matching}\\
Matching errors come from the $m_{\perp,\parallel}$ fit parameters as explained 
in the previous section.
\end{itemize}
Physical meson mass input errors (0.01\%) and finite size errors (0.1\%) 
are not included in the plots, since they are too small to have any effect.

\vspace{.1in}
The slope of $f_+(q^2)$ as one comes down from the zero recoil 
point at $q^2 = q^2_{max}$ is a quantity that is often quoted when comparing 
different measurements of this form factor. In terms of the variable $w = (M_B^2 + M_D^2 - q^2)
/(2 M_B M_D)$ 
the slope parameter $\rho^2$ is given by
\be
{\cal G}(w) =  {\cal G}(1) \; \left \{ 1 - \rho^2 (w - 1) + {\cal O}((w-1)^2) \right \},
\ee
where,
\be
{\cal G}(w=w(q^2)) = \frac{2 \sqrt{\kappa}}{1+\kappa} f_+(q^2)
\ee
for
\be
\kappa = \frac{M_D}{M_B}.
\ee

\vspace{.1in}
A popular way to extract $\rho^2$ is to use the Caprini-Lellouch-Neubert (CLN) 
parametrization \cite{cln},
\begin{eqnarray}
\label{cln}
{\cal G}(w) & = & {\cal G}(1) \; \left \{ 1 - 8 \rho^2 z + (51 \rho^2 - 10) z^2 \right . 
\nonumber \\
 & &  \left . \qquad \qquad  \quad - (252 \rho^2 - 84 ) z^3  \right \},
\end{eqnarray}
with,
\be
z = \frac{\sqrt{w+1} - \sqrt{2}}{\sqrt{w+1} + \sqrt{2}}.
\ee 
This ``$z$'' is the same as the z-variable introduced in the previous 
section, Eq.~(\ref{zvar}), with the same $t_0 = q^2_{max}$.   Using Eq.~(\ref{cln}), 
we extract
\be
\rho^2 = 1.119(71), \qquad \qquad {\cal G}(1) = 1.035(40).
\ee
Another useful reference point is the value of $f_+(0) = f_0(0)$.  We find,
\be
f_+(0) = 0.664(34).
\ee
In Appendix A we provide the z-expansion coefficients including errors
and  correlations for the form factors of Fig.~\ref{f0fp}.

\section{Extraction of $|V_{cb}|$}

The differential branching fraction for $B \rightarrow D l \nu$ decays is 
given by,
\begin{eqnarray}
\label{diffbr}
& & \frac{d \Gamma}{d q^2} = \eta_{EW} \frac{G_F^2 |V_{cb}|^2}{24 \pi^3 M_B^2} (1-\frac{m_l^2}{q^2})^2 |\vec{p}| \\
&\times& \left [ (1+\frac{m_l^2}{2q^2}) M_B^2 |\vec{p}|^2 f_+^2(q^2) + \frac{3 m_l^2}{8 q^2} (M_B^2-M_D^2)^2 f_0^2(q^2) \right ], \nonumber
\end{eqnarray}
where $m_l$ is the mass of the lepton, and $\eta_{EW}$ is the electro-weak correction. 
The main goal of the present work is to combine experimental measurement of 
this differential branching fraction with form factors of the previous section 
to extract $|V_{cb}|$.  
The partial branching fraction (the left hand side of Eq.~\ref{diffbr}) 
has been measured by BABAR \cite{babar2010}. 
On the right hand side, we have form factors from this lattice calculation, and 
 all other factors are known except the target quantity $|V_{cb}|$.

\vspace{.1in}
In order to include the higher order electro-weak effects, we apply the Sirlin 
factor~\cite{sirlin}, $\eta_{s} = 1.00662$.  Furthermore, there are final state 
electro-magnetic interactions for the neutral channel, $\bar{B}^0 \rightarrow D^+ l \nu$,
which we estimate to be a less than 0.5\%  effect using the signal yield ratio of
 the charged and neutral decay channels. Combining the two effects, 
we get $\eta_{EW}  = 1.011(5)$.

\begin{figure}
\includegraphics*[width=10.0cm,height=8.0cm,angle=0]{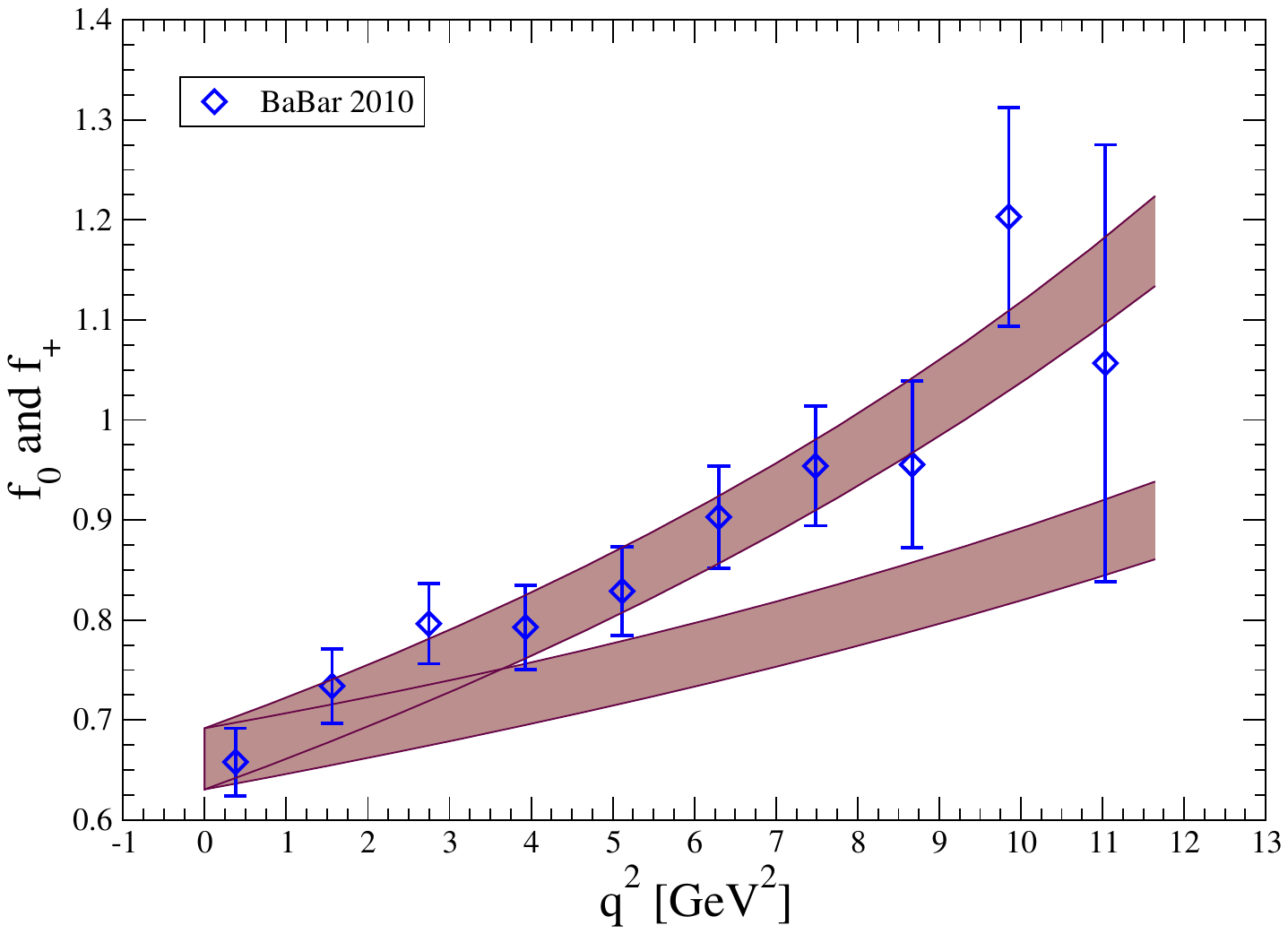}
\caption{
Form factors using both lattice and BABAR ~\cite{babar2010} inputs, together with 
the experimental data points. }
\label{withExp}
\end{figure}

\vspace{.1in}
We perform another modified z-expansion fit explained in Sec. IV together with 
the BABAR experiment data with $|V_{cb}|$ as a fit parameter. 
 We have a good fit with $\chi^2/dof = 0.88$, and this is shown in Fig.~\ref{withExp}.
 We get $|V_{cb}|$ from this fit,
\be
\label{vcb2}
|V_{cb}| = 0.0402(17)(13),
\ee 
where the first error is from the fit including all lattice errors and experimental
 statistical errors, and the second error is the experimental systematic error.
We quote the experimental systematic errors as 3.3\% of our fit result based on 
 BABAR's estimate of  their systematic errors in~\cite{babar2010}.
  This is equivalent to imposing 3.3\% systematic errors on each experimental
  measurement bin with 100\% correlations.  

\vspace{.1in}
A detailed error budget is shown in Table~V.
The dominant errors are experimental systematic, lattice discretization, and 
operator matching errors.
Thus, improvements in both experiments and lattice calculations are required to obtain 
better precision on $|V_{cb}|$ from our method.

\begin{figure}
\includegraphics*[width=10.0cm,height=8.0cm,angle=0]{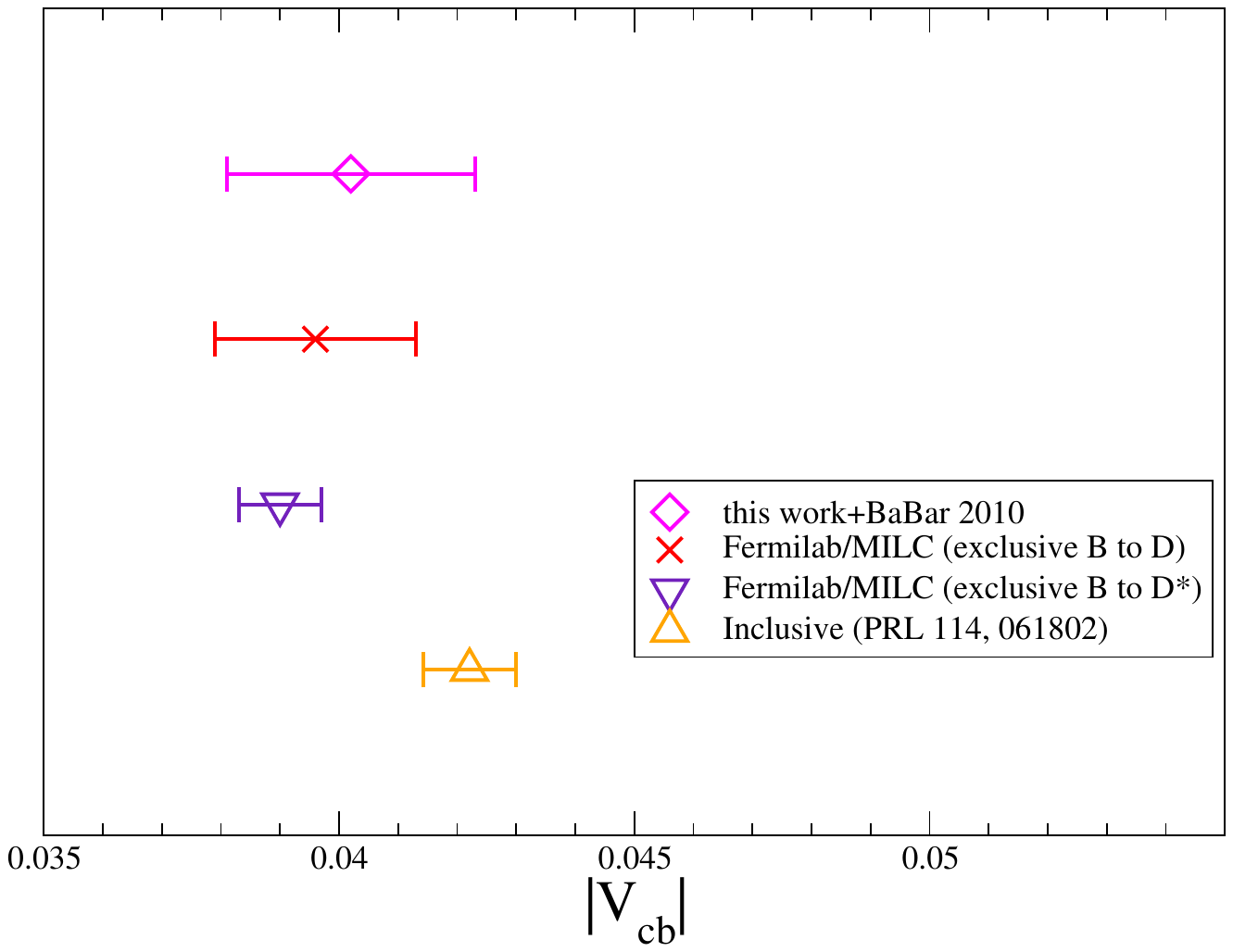}
\caption{
$|V_{cb}|$ comparisons between inclusive and exclusive determinations. }
\label{vcb}
\end{figure}

\vspace{.1in}
$|V_{cb}|$ has been reported from multiple lattice and non-lattice calculations.
We compare the different determinations in Fig.~\ref{vcb}. 
Our result agrees with other exclusive calculations, particularly with the most
 accurate result from $B \rightarrow D^* l \nu$, but it is also compatible within errors 
 with the inclusive determination.  Since the discretization error is one of the dominant
 errors in our calculation, lattice errors can be reduced in the future by 
working on more ensembles with finer lattice spacings.

\begin{table}
\caption{
Error budget table for $|V_{cb}|$. The first three rows are from experiments, and the rest are from lattice simulations. 
}
\begin{ruledtabular}
\begin{tabular}{cc}
Type   & Partial errors [\%]    \\
\hline
experimental statistics  & 1.55  \\
experimental systematic   &  3.3  \\
meson masses   & 0.01  \\
\hline
lattice statistics  &  1.22  \\
chiral extrapolation   & 1.14  \\
discretization   & 2.59  \\
kinematic   & 0.96  \\
matching   & 2.11  \\
electro-weak   & 0.48 \\
finite size effect   & 0.1  \\
\hline
total  &  5.34  \\
\end{tabular}
\end{ruledtabular}
\end{table}

\section{The R(D) Ratio}
The experimental data used in the previous section to extract $|V_{cb}|$ were for 
semileptonic decays with light leptons in the final state.  
BABAR has also studied decays involving the much heavier $\tau$ lepton, 
$ B \rightarrow D \tau \nu_\tau$, and measured the ratio,
\be
R(D) = \frac{{\cal B}(B \rightarrow D \tau \nu_\tau)}{{\cal B}(B \rightarrow D l \nu)},
\ee
where $l$ is either an electron or a  muon.
They find 
\be
R(D)|_{exp.} = 0.440 (58) (42),
\ee
where the first error is the statistical and the second  is the 
systematic error~\cite{rdbabar}.

\begin{figure}
\includegraphics*[width=10.0cm,height=8.0cm,angle=0]{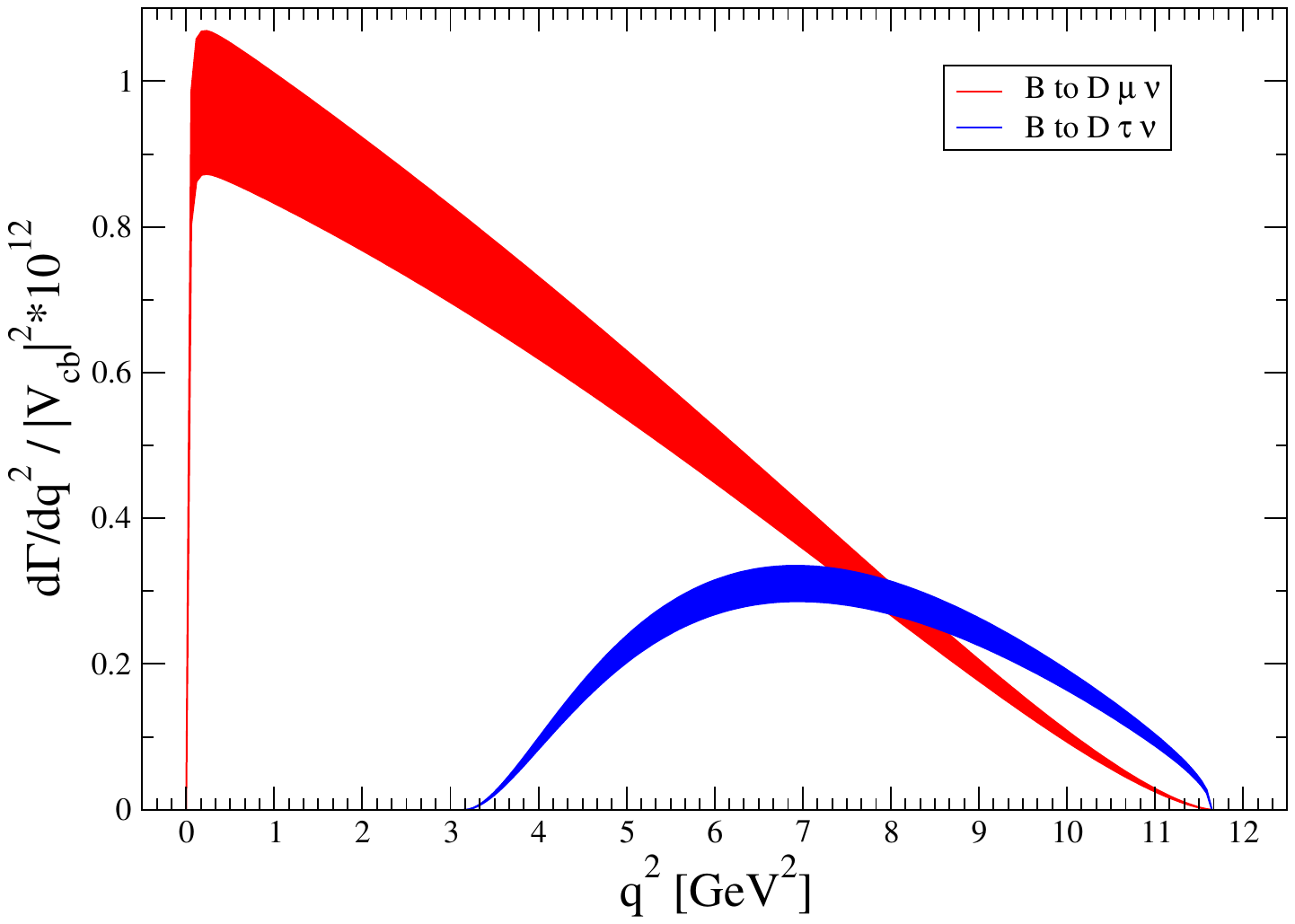}
\caption{
The differential branching fractions for $B \rightarrow D l \nu$ and $B \rightarrow D \tau \nu$ }
\label{pbr}
\end{figure}

\begin{figure}
\includegraphics*[width=10.0cm,height=8.0cm,angle=0]{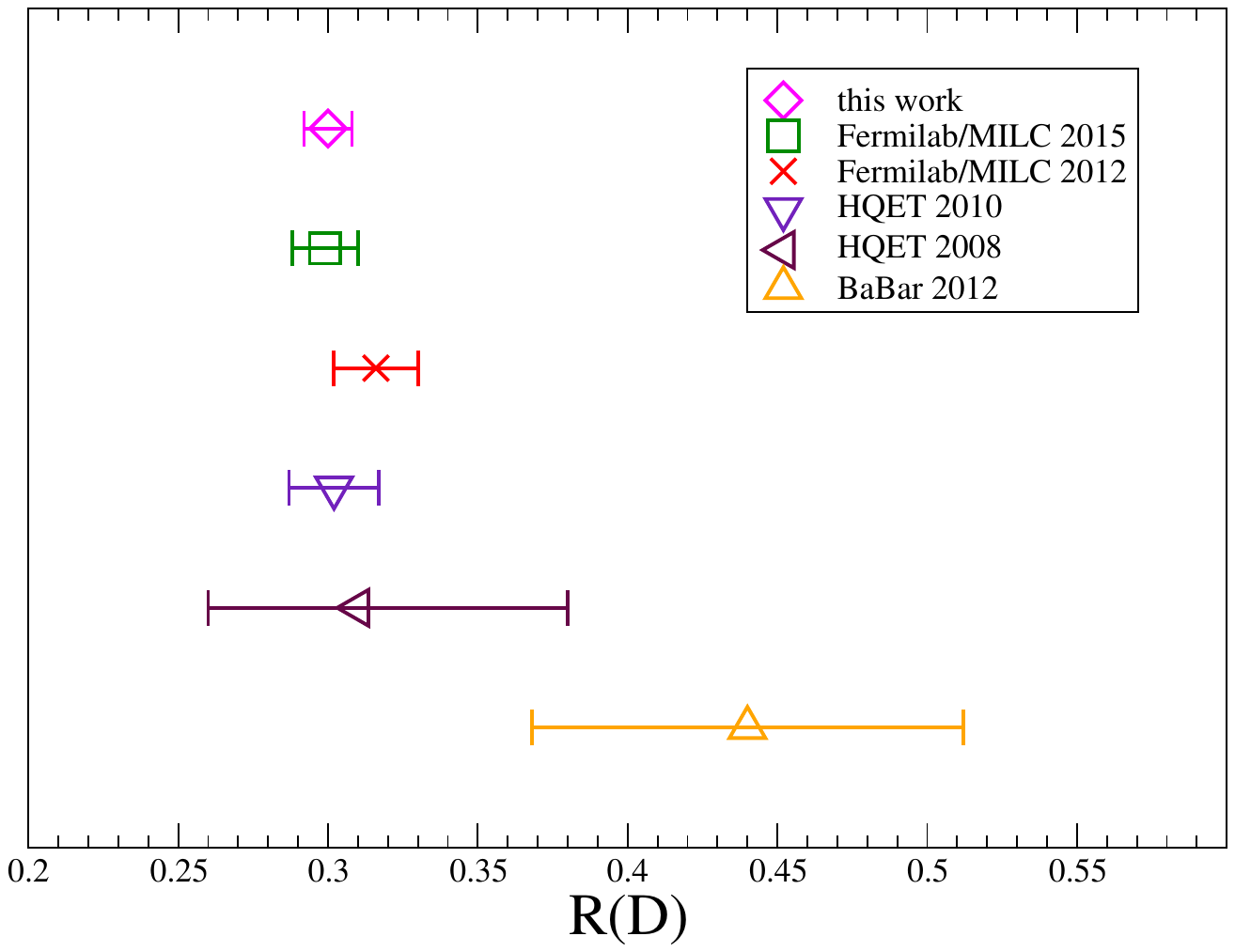}
\caption{
Comparisons between different determinations of $R(D)$. The references for the other determinations are BABAR 2012~\cite{rdbabar}, HQET 2008~\cite{rdhqet2}, HQET 2010~\cite{rdhqet}, Fermilab/MILC 2012~\cite{RdMILC}, and Fermilab/MILC 2015~\cite{DMILC}.}
\label{rd}
\end{figure}

\vspace{.1in}
Here we present a Standard Model prediction for $R(D)$ based on 
 our new form factors. Fig.~\ref{pbr} compares differential branching fractions 
of Eq.~(\ref{diffbr}) for $B \rightarrow D \tau \nu_\tau$  and for 
$B \rightarrow D l \nu$.  Although only $f_+(q^2)$ 
contributes to the $l \nu$ case, both $f_+(q^2)$ and $f_0(q^2)$ are involved 
in the $\tau \nu_\tau$ branching fraction.  Integrating over $q^2$ we obtain,
\be
\label{rd2}
R(D) |_{SM} = 0.300(8).
\ee
Table VI shows a detailed error budget for $R(D)$.
Fig.~\ref{rd} gives a comparison plot for different determinations of $R(D)$.
All Standard Model based calculations are in good agreement with each other. The difference 
between our result and experiment is at the $2\sigma$ level.
 We note that we do not use any experimental results to extract $R(D)$. 
 Our result gives the most accurate pure Standard Model prediction to date for $R(D)$. 

\begin{table}
\caption{
Error budget table for $R(D)$.
}
\begin{ruledtabular}
\begin{tabular}{cc}
Type   & Partial errors [\%]    \\
\hline
lattice statistics  &  1.24  \\
chiral extrapolation   & 0.28  \\
discretization   & 1.08  \\
kinematic   & 1.61  \\
matching   & 1.03  \\
finite size effect   & 0.1  \\
\hline
total  &  2.54  \\
\end{tabular}
\end{ruledtabular}
\end{table}

\section{Summary and Future Prospects}

In this paper we have presented a new lattice QCD calculation of the 
$B \rightarrow D l \nu$ semileptonic decay form factors $f_+(q^2)$ and 
 $f_0(q^2)$. These were  combined with experimental measurements of 
differential branching fractions to extract a value for $|V_{cb}|_{excl.}$. 
Our result, given in Eq.~(\ref{vcb1}) (and repeated in (\ref{vcb2})) is 
consistent with other recent lattice determinations using 
different lattice actions, and provides a cross check of earlier calculations. 
We summarize these results in Fig.~\ref{vcb}.

\vspace{.1in}
 The dominant 
error in our calculation is the discretization error, followed by higher 
order current matching uncertainties. 
The former error can be reduced by adding simulation data from further ensembles with finer 
lattice spacings.  We are also exploring ways to improve our matching errors by 
combining simulations with NRQCD bottom-quarks  with those employing  
heavier than charm HISQ quarks.  This approach to nonperturbative matchings 
of NRQCD/HISQ currents is described briefly in the Appendix to Ref. 
\cite{bstok}.  There we presented ratios of $B_s \rightarrow K l \nu$ 
and $B_s \rightarrow \eta_s l \nu$ form factors and explained how such ratios 
combined with a purely HISQ calculation in the future  of $B_s \rightarrow \eta_s l \nu$ 
form factors  would lead to a nonperturbative determination 
of the NRQCD/HISQ 
bottom-up current Z-factors.  Similarly, nonperturbative Z-factors for 
bottom-charm currents used in the present calculation 
could be obtained by calculating $B_s \rightarrow D_s l \nu$ 
forms factors once with NRQCD bottom-quarks and then again with heavy-HISQ 
bottom-quarks and then taking ratios.  We have already completed, and are in 
the process of writing up, calculations of $B_s \rightarrow D_s l \nu$ form factors 
with NRQCD bottom-quarks.  Simulations with heavy-HISQ bottom-quarks are also 
underway.  Hence we expect to be able to significantly reduce theory 
errors in $|V_{cb}|$ determinations from $B \rightarrow D l \nu$ decays 
in the near future. In the meantime we hope that experimental measurements 
will also improve considerably. Only then will one be able to shed light 
on the exclusive versus inclusive tensions for $|V_{cb}|$ via studies of 
$B \rightarrow D l \nu$ decays.

\vspace{.1in}
In this article we also determined the ratio $R(D)$. 
Our result is given in Eq.~(\ref{rd0})  (and again in (\ref{rd2})).  We summarize comparisons between 
Standard Model predictions and experiment in Fig.~\ref{rd}.  It will be 
interesting to see whether the current $\sim$2$\sigma$ tension will develop 
into a true discrepancy between experiment and the Standard Model 
 or disappear.

\vspace{.2in}
{\bf Acknowledgements}: \\
H. N. is supported in part by the U.S. Department of Energy (DOE) under Grant No. DE-FC02-12ER41879 and by
the U.S. National Science Foundation (NSF) under Grant No. PHY10-034278; C. M. B. and J. S. by U.S. DOE under 
Grant No. DE-SC0011726 and C. M. by U.S. NSF under Grant No. PHY10-034278. Numerical simulations were
 carried out on facilities
of the USQCD collaboration funded by the Office of Science of the DOE and 
at the Ohio Supercomputer Center.
We thank the MILC collaboration for use of their gauge configurations.

\appendix
\section{Reconstructing  Form Factors}
We provide our z-expansion coefficients with correlations, so that readers can reconstruct our form factors for their analysis.
The form factors are expressed by the BCL parameterization as
\be
f_+(q^2) = \frac{1}{P_+} \sum_{k=0}^{2} a_k^{(+)}[ z(q^2)^k - (-1)^{k-K}\frac{k}{K}z(q^2)^K ],
\ee
and
\be
f_0(q^2) = \frac{1}{P_0}\sum_{k=0}^{2} a_k^{(0)} z(q^2)^k,
\ee
where 
\begin{eqnarray}
z(q^2) &=& \frac{\sqrt{t_+ - q^2} - \sqrt{t_+ - t_0}}{ \sqrt{t_+ - q^2} + \sqrt{t_+ - t_0} }, \\
t_+ &=& (M_B + M_D)^2, \\
t_0 &=& q^2_{max} = (M_B - M_D)^2, \\
P_{+,0}(q^2) &=& \left ( 1 - \frac{q^2}{M_{+,0}^2} \right ).
\end{eqnarray}
For the locations of the poles, one can use $M_+ = 6.330(9)$ GeV for $f_+$, and $M_0 = 6.420(9)$ GeV for $f_0$ to reproduce our form factors exactly.
The coefficients, $a_k^{(+,0)}$, and the correlations are presented in Table~VII.

\begin{table}
\caption{
z-expansion coefficients and its covariance.
}
\begin{ruledtabular}
\begin{tabular}{cc}
coefficient   & value  \\
\hline
$a_0^{(0)}$ &   0.647(29) \\
$a_1^{(0)}$ &    0.27(30) \\
$a_2^{(0)}$ &    -0.09(2.94)\\
$a_0^{(+)}$ &  0.836(33)  \\
$a_1^{(+)}$ &   -2.66(52) \\
$a_2^{(+)}$ &   -0.07(2.96) \\
\hline
\end{tabular}
\begin{tabular}{ccccccc}
  &  $a_0^{(0)}$ &$a_1^{(0)}$ & $a_2^{(0)}$ & $a_0^{(+)}$ & $a_1^{(+)}$ & $a_2^{(+)}$ \\
\hline
$a_0^{(0)}$ &  8.442e-4 & -1.141e-3 & -5.072e-3 &  4.799e-4
   & 3.801e-3  & 5.518e-3     \\
$a_1^{(0)}$ &   &    9.255e-2 & -1.087e-1  & 5.390e-4
   & 5.835e-2  & 1.852e-2  \\
$a_2^{(0)}$ &     &    &   8.652 &   6.813e-3
   & 2.504e-1 &  2.402e-1    \\
$a_0^{(+)}$ &    &    &    &    1.062e-3
   &-7.548e-3 & -7.354e-3  \\
$a_1^{(+)}$ &     &    &    &    &     2.747e-01 & -3.561e-1    \\
$a_2^{(+)}$ &     &    &    &    &     &   8.740  \\
\end{tabular}
\end{ruledtabular}
\end{table}

\section{Chiral/Continuum Extrapolations using Input from HPChPT}
\begin{figure}
\includegraphics*[width=10.0cm,height=8.0cm,angle=0]{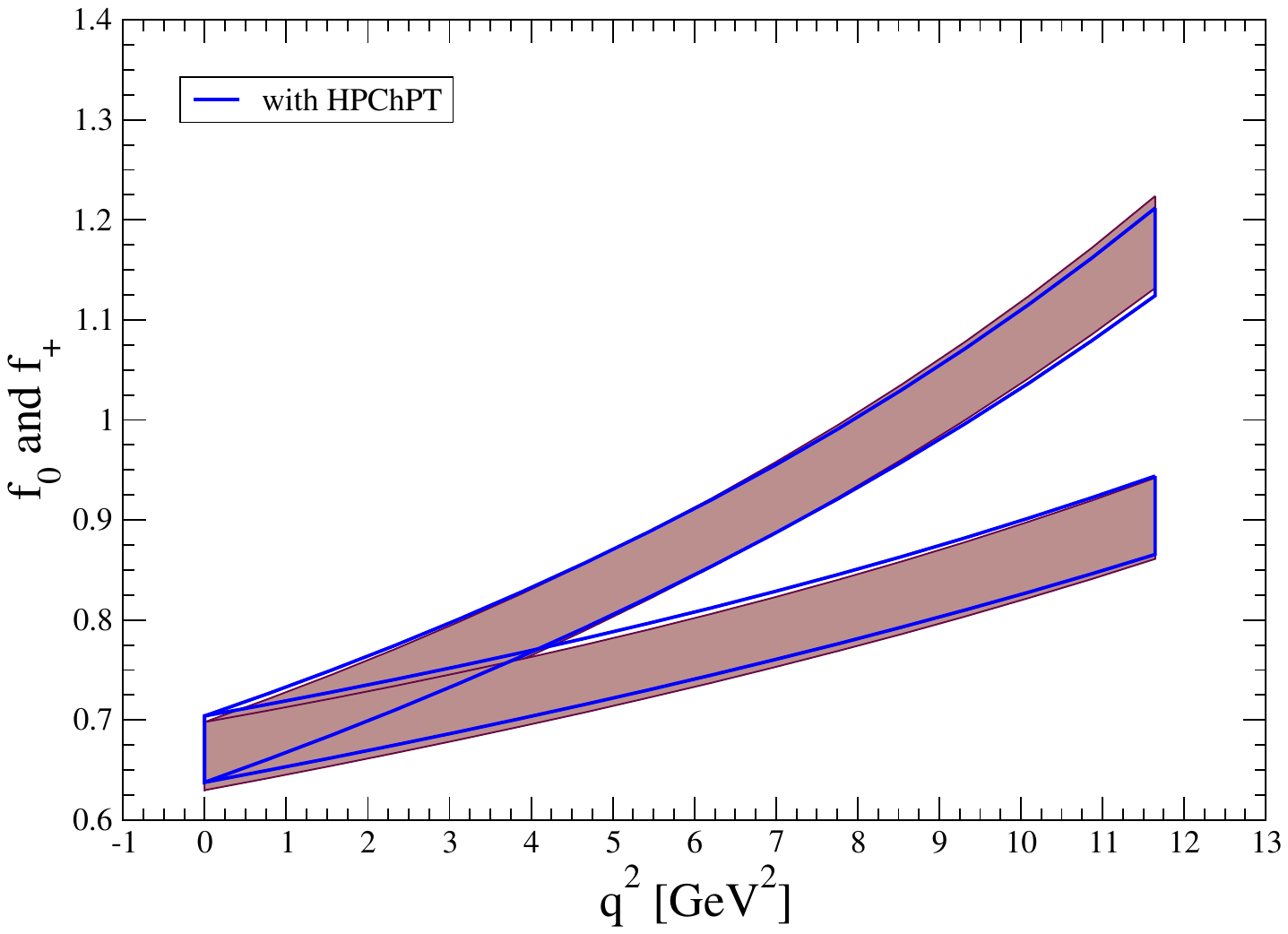}
\caption{
Comparison between using a generic $x_\pi {\rm log}(x_\pi)$ term (filled blocks) and 
chiral logarithms from HPChPT (open blocks).}
\label{chlog}
\end{figure}

In the standard extrapolation of Sec. IV we used a generic $c^k_3 x_\pi 
{\rm log}(x_\pi)$ term to parametrize chiral logarithmic contributions 
and allowed $c^k_3$ to float. 
An alternate way to introduce chiral logarithms into our chiral/continuum 
extrapolations is to use expressions fixed by 
 hard pion chiral perturbation theory (HPChPT) \cite{hpchpt},
\begin{eqnarray}
\label{logs1}
[\text{logs}]_{f_+} &=& -\, \frac{\kappa+1}{\sqrt{\kappa}} 
\frac{g^2}{(4 \pi f_\pi)^2}
  (r(w)-1) \nonumber \\ 
& & \times \;  \left (\frac{3}{2}\bar{A}(x_\pi) + \bar{A}(x_K) + \frac{1}{6} 
\bar{A}(x_\eta) \right ),
\end{eqnarray}
\begin{eqnarray}
\label{logs2}
[\text{logs}]_{f_0} &=& - \frac{\sqrt{\kappa}}{1+\kappa} \frac{g^2}{(4 \pi f_\pi)^2}
 (w + 1) \,(r(w)-1) \nonumber \\ 
& & \times \;    \left (\frac{3}{2}\bar{A}(x_\pi) + \bar{A}(x_K) + \frac{1}{6} 
\bar{A}(x_\eta) \right ),
\end{eqnarray}
 where
 \begin{eqnarray}
\bar{A}(x) &=& x \, {\rm log}(x), \\
 w &=& \frac{M_B^2 +M_D^2 - q^2}{2M_BM_D}, \\
 r(w) &=&  \frac{1}{\sqrt{w^2-1}}{\rm log}(w+\sqrt{w^2-1}), \\
 \kappa &=& M_D/M_B.
 \end{eqnarray}
We find very consistent results for the extrapolated physical form factors 
using either generic $c^k_3$ terms or (\ref{logs1}) \& (\ref{logs2}). 
This is already evident in Fig.~\ref{tests} test number 10 for 
$f_+(0)$ and $f_+(q^2_{max})$. In Fig.~\ref{chlog} we compare the two 
approaches over the entire $q^2$ range.

\section{Priors and Prior Widths for the 
Chiral/Continuum/Kinematic Extrapolation}
In earlier works~\cite{dtok,dtopi}, we split the priors for the modified z-expansion method into two groups: Group I and Group II.
The Group I parameters are typical fit parameters, such as quark mass dependence or z-expansion parameters.  In this work, the Group I parameters consist of
\be
c_1^k, c_2^k, c_3^k, d_1^k, d_2^k, e_1^k, e_2^k, a_k,
\ee 
where $k=0,1,$ and $2$, and there are two sets of parameters for each $f_0$ and $f_+$ form factors.  These parameters are defined in Eq.~(\ref{bcl1}),~(\ref{bcl2}), and~(\ref{dk}), and the priors and fit results are shown in Table VIII. 

\vspace{.1in}
We choose priors as following. For the valence quark mass terms, $c_1^k$, we use $0.0(1.0)$, since the mass terms are normalized by the scale, $4\pi f_\pi$. However, it is well-known that the sea quark mass effects are smaller than those of the valence quark effects, so we take $0.0(3)$ for the sea quark mass terms, $c_2^k$.
HPChPT suggests a prior for the generic chiral log term, $x_\pi log(x_\pi)$, as   $0.0(1)$.  This prior essentially covers variations of the terms on entire kinematic range.  For more conservative error estimations, we take $0.0(2)$ as our prior for the generic chiral log term.  We note that the prior settings with $0.0(1)$ and $0.0(2)$ give almost identical results.   
In the HISQ action, leading heavy quark discretization errors are ${\cal O}(\alpha_s v^2/c^2 am_c^2)$ and ${\cal O}( v^2/c^2 am_c^4)$.  We conservatively do not take the $v^2/c^2$ terms in our power counting, so that we take $0.0(3)$ for $d_1^k$ and $e_1^k$ priors, and $0.0(1.0)$ for $d_2^k$ and $e_2^k$ priors.
For the priors for z-expansion coefficients, $a_k^{(+,0)}$, we searched for broad enough priors that gave stable fit results, and we take $0.0(3.0)$ in this work.

\begin{table}[t]
\caption{
Priors and fit results of the Group I parameters for the modified z-expansion fit. 
}
\begin{ruledtabular}
\begin{tabular}{ccccc}
Group I   & prior [$f_0$] & fit result [$f_0$] &   prior [$f_+$] & fit result [$f_+$] \\
\hline
$c_1^0$ &  0.0 (1.0)  &  -0.09 (18) & 0.0 (1.0)  &  0.34 (20)  \\
$c_1^1$ &  0.0 (1.0)  &  -0.13 (99) &  0.0 (1.0) &  -0.67 (87)  \\
$c_1^2$ &  0.0 (1.0)  & 0.0 (1.0)  &  0.0 (1.0) &   0.0 (1.0) \\
\hline
$c_2^0$ &  0.00 (30)  &  0.03 (28)  & 0.00 (30)  &  -0.10 (28)  \\
$c_2^1$ & 0.00 (30)   & 0.00 (30)  & 0.00 (30)  &   -0.01 (30) \\
$c_2^2$ & 0.00 (30)   & 0.00 (30)  &  0.00 (30) &  0.00 (30)  \\
\hline
$c_3^0$ &  0.00 (20)  & -0.10 (15)  & 0.00 (20)  & 0.22 (16)   \\
$c_3^1$ &   0.00 (20) &  0.006 (200) & 0.00 (20)  &  0.03 (20)  \\
$c_3^2$ & 0.00 (20)   &  -0.00 (20) &  0.00 (20) &  0.00 (20)  \\
\hline
$d_1^0$ &  0.00 (30)  & -0.16 (24)  & 0.00 (30)  &  0.11 (24)  \\
$d_1^1$ &  0.00 (30)  &  0.02 (30) &  0.00 (30) &   -0.005 (292) \\
$d_1^2$ &  0.00 (30)  & -0.00 (30)  & 0.00 (30)  &   0.00 (30) \\
\hline
$d_2^0$ &  0.0 (1.0)  &  -0.17 (44) & 0.0 (1.0)  &  -0.29 (40)  \\
$d_2^1$ &  0.0 (1.0)  & 0.2 (1.0)  &  0.0 (1.0) &   0.008 (923) \\
$d_2^2$ &  0.0 (1.0)  &  -0.0 (1.0) & 0.0 (1.0)  &  0.0 (1.0)  \\
\hline
$e_1^0$ &  0.00 (30)  &  0.21 (25) & 0.00 (30)  &  0.06 (25)  \\
$e_1^1$ &  0.00 (30)  & 0.008 (300)  &  0.00 (30) &  -0.005 (298)  \\
$e_1^2$ &  0.00 (30)  &  0.00 (30) & 0.00 (30)  &  0.00 (30)  \\
\hline
$e_2^0$ & 0.0 (1.0)   &  1.44 (66) & 0.0 (1.0)  &  0.03 (82)  \\
$e_2^1$ & 0.0 (1.0)   &  0.02 (1.00) &  0.0 (1.0) &   0.0 (1.0) \\
$e_2^2$ & 0.0 (1.0)   & 0.0 (1.0)  & 0.0 (1.0)  &  0.0 (1.0)  \\
\hline
$a_0$ & 0.0 (3.0)   &  0.644 (30) & 0.0 (3.0)  &  0.842 (35)  \\
$a_1$ &  0.0 (3.0)  & 0.27 (31)  &  0.0 (3.0) &  -2.69 (54)  \\
$a_2$ &  0.0 (3.0)  &  -0.09 (2.94) & 0.0 (3.0)  &  -0.07 (2.96)  \\
\end{tabular}
\end{ruledtabular}
\end{table}

\vspace{.2in}
The Group II parameters are
\begin{eqnarray}
& \left ( \frac{r_1}{a} \right )^i, aM_B^i, aE_D^i(\vec{p}), aM_\pi^i, (aM_K^{asqtad})^i, (aM_\pi^{asqtad})^i,\nonumber \\
& M_0, M_+, r_1, M_\pi^{phys}, M_K^{phys}, M_B^{phys}, M_D^{phys},
\end{eqnarray}
where $i$ is the index for the five ensembles ($i = 1,2,3,4,$ and $5$). 
The Group II parameters are either from experiments, from other lattice simulations, or from the correlator fits, and used for input parameters.  The prior settings and fit results for the Group II are shown in Table IX and X.

\begin{table}[t]
\caption{
Priors and fit results of the Group II parameters for the modified z-expansion fit.  Parameters with five rows are lattice quantities for the five ensembles, C1, C2, C3, F1, and F2. 
}
\begin{ruledtabular}
\begin{tabular}{ccccc}
Group II   & prior  & fit result  \\
\hline
$r_1/a$ &  2.6470 (30)  &   2.6473 (30)  \\
   &  2.6180 (30)  &   2.6174 (30)  \\
   &  2.6440 (30)  &   2.6442 (30)  \\
   &   3.6990 (30) &  3.6991 (30)   \\
   &   3.7120 (40) &  3.7120 (39)   \\
 $aM_B$  &  3.18915 (65)  &   3.18905 (65)  \\
   &  3.23184 (88)  &   3.23197 (87)  \\
   &  3.21191 (77)  &  3.21177 (77)   \\
   &  2.28109 (52)  &   2.28092 (50)  \\
   &  2.28101 (44)  &   2.28105 (44)  \\
$aE_D$   &  1.1389 (10)  &    1.13894 (82)  \\
 $\vec{p}=(0,0,0)$  &   1.15993 (82)  &  1.16011 (80)   \\
   &  1.16339 (54)  &  1.16333 (54)   \\
   &  0.81452 (35)  &   0.81444 (35)  \\
   &   0.81993 (27) &   0.81997 (27)  \\
$aE_D$   & 1.1688 (11)   &  1.16827 (88)   \\
 $\vec{p}=(1,0,0)$  &  1.19901 (99)  &   1.19918 (94)  \\
   & 1.20395 (77)   &  1.20445 (69)   \\
   &  0.84360 (58)  &   0.84374 (52)  \\
   &  0.85071 (40)  &   0.85055 (39)  \\
 $aE_D$  &  1.19884 (84)  &   1.19847 (80)  \\
  $\vec{p}=(1,1,0)$ &  1.24003 (87)  &  1.23982 (83)   \\
   &  1.24485 (78)  &   1.24477 (71)  \\
   &  0.87300 (62)  &  0.87299 (57)   \\
   &  0.87885 (36)  &  0.87890 (35)   \\
$aE_D$   &  1.22775 (96)  &  1.22746 (94)   \\
$\vec{p}=(1,1,1)$   &  1.27839 (93)  &  1.27825 (91)   \\
   & 1.28321 (94)   &   1.28319 (89)  \\
   &  0.90004 (78)  &  0.90027 (70)   \\
   &  0.90627 (49)  &   0.90630 (46)  \\
\end{tabular}
\end{ruledtabular}
\end{table}


\begin{table}[b]
\caption{
(Continued) Priors and fit results of the Group II parameters for the modified z-expansion fit.  Parameters with five rows are lattice quantities for the five ensembles, C1, C2, C3, F1, and F2. 
}
\begin{ruledtabular}
\begin{tabular}{ccccc}
Group II   & prior  & fit result  \\
\hline
$aM_\pi$  &  0.15990 (20)  &   0.15990 (20)  \\
   &  0.21110 (20)  &   0.21110 (20)  \\
   &  0.29310 (20)  &  0.29310 (20)   \\
   &  0.13460 (10)  &   0.13460 (10)  \\
   &   0.18730 (10) &   0.18730 (10)  \\
 $aM_K^{asqtad}$  &   0.36530 (29) &  0.36530 (29)   \\
   &   0.38331 (24)  &  0.38331 (24)   \\
   & 0.40984 (21)   &  0.40984 (21)   \\
   &  0.25318 (19)  &  0.25318 (19)   \\
   &  0.27217 (21)  &  0.27217 (21)   \\
 $aM_\pi^{asqtad}$  &   0.15971 (20) &   0.15971 (20)  \\
   & 0.22447 (17)   &  0.22447 (17)   \\
   &  0.31125 (16)  &  0.31125 (16)   \\
      & 0.14789 (18)   & 0.14789 (18)    \\
   &  0.20635 (18)  &   0.20635 (18)  \\
  $M_0$ & 6.53 (1.00)   &  6.42(43)   \\
  $M_+$ &  6.3300 (90)  &  6.3300(90)   \\
    $r_1$  &  0.3133 (23)  &  0.3132 (23)   \\
  $M_\pi^{phys}$ &  0.1373 (23)  &  0.1373 (23)   \\
  $M_K^{phys}$ &  0.4957 (20)  &   0.4957 (20)  \\
    $M_B^{phys}$  &   5.27942 (17)  &   5.27942 (17)  \\
  $M_D^{phys}$ & 1.86690 (40)   &   1.86690 (40)  \\
\end{tabular}
\end{ruledtabular}
\end{table}




\end{document}